\documentclass[manuscript]{aastex}

\usepackage{lscape}

\usepackage[usenames,dvips]{color}
\usepackage{epsfig}
\usepackage{graphicx}

\def\fermi{{\it Fermi}}

\shorttitle{High-energy emissions from LS~5038}
\shortauthors{Takata et al.}

\begin{document}


\title{High-energy emissions from the gamma-ray binary LS~5039}

 \author{J. \textsc{Takata}\altaffilmark{1},
Gene C.K.\textsc{Leung}\altaffilmark{1},
P.H.T. \textsc{Tam}\altaffilmark{2},
A.K.H.  \textsc{Kong}\altaffilmark{2},
C.Y. \textsc{Hui}\altaffilmark{3},
and 
K.S. \textsc{Cheng}\altaffilmark{1}
   }
\altaffiltext{1}{Department of Physics, University of Hong Kong, Pokfulam Road, Hong Kong}
\altaffiltext{2}{Institute of Astronomy and Department of Physics, National 
Tsing Hua University, Hsinchu, Taiwan}
\altaffiltext{3}{Department of Astronomy and Space Science, Chungnam National
University, Daejeon, Republic of Korea}
\email{takata@hku.hk,  gene930@connect.hku.hk,  hrspksc@hku.hk}


\begin{abstract}
We study mechanisms of multi-wavelength  emissions (X-ray, GeV and TeV 
gamma-rays) from the gamma-ray binary LS~5039. This paper is composed of  two 
parts. In the first part, we report on results of observational analysis using 
 four year data of \fermi\ Large Area Telescope.  
 Due to the improvement of instrumental response function 
 and increase of the statistics, the observational 
uncertainties of the spectrum in  
 $\sim$100-300 MeV bands and $>10$GeV bands are  significantly improved. 
 The present data analysis suggests that the 0.1-100GeV emissions 
from LS~5039 contain three different components; (i) the first 
 component contributes to $<$1GeV emissions around superior conjunction, (ii) the second component dominates in 1-10GeV energy bands and (iii) the third
 component is compatible to lower energy tail of the TeV emissions.
In the second part, we develop  an emission model to 
 explain the properties of the phase-resolved emissions in 
multi-wavelength observations. Assuming that LS~5039 includes a pulsar, 
we argue that both emissions from 
  magnetospheric outer gap  and inverse-Compton scattering 
 process of  cold-relativistic pulsar wind contribute to the observed 
GeV emissions. We assume that the pulsar is wrapped by 
 two kinds of termination 
shock; Shock-I due to the interaction between 
the pulsar wind and the stellar wind  and Shock-II due to the 
effect of the orbital motion. We propose that the  X-rays are produced by 
the synchrotron radiation at Shock-I region and the TeV gamma-rays 
are produced by  the inverse-Compton scattering process 
at Shock-II region.
\end{abstract}





\section{Introduction}
\label{intro}

The gamma-ray binary is a class  binary  system
emitting  high-energy (GeV and/or TeV) gamma-rays, 
and comprises  a compact object (neutron star or black hole) and
  a high-mass OB star (see Dubus 2013 for recent review on the 
gamma-ray binaries). Their radiation spectra have   a peak in $\nu F_{\nu}$
 around GeV energy bands, and extends up to  $1-10$TeV energy bands. 
 Five gamma-ray binaries have been detected so far, namely,  
PSR B1259-63/LS2883 system (Aharonian et al. 2005),
 LS~5039 (Aharonian et al. 2006), LS~I$\mathrm{+61^{o}}$~303 
(Albert et al. 2006),  1FGL~J1018.6-5856 (Ackermann et al. 2012) and 
H.E.S.S.~J0632+057 (Hinton et al. 2009). PSR B1259-63/LS~2883 
is the only binary system for which the compact object has been 
confirmed to be a young pulsar. 

The GeV gamma-ray observation of the $Fermi$  telescope provides 
a new challenge for understanding of the non-thermal emission process 
around  gamma-ray binary. The \fermi\ has revealed that different
 systems show different properties of the GeV emissions. 
 PSR B1259-63/LS~2883 showed
 a weak and flare-like  emissions during the 2010-2011 periastron 
passage (Abdo et al. 2011; Tam et al. 2011). The GeV emissions from LS~5039, 
 LS~I$\mathrm{+61^{o}}$~303 and  1FGL~J1018.6-5856 are observed 
for entire orbit, and the spectra are fitted by a power-law plus 
 exponential cut-off form with a cut-off energy  around several GeV.  
The emissions from LS~I$\mathrm{+61^{o}}$~303 show a long-term
 variability related with the 1667 day super-orbital period in 
radio (Ackermann et al. 2013).  No detection of the GeV emissions 
has been reported for H.E.S.S.~J0632+057. 

The gamma-ray binary LS~5039 has been  known 
as a ralatively  compact binary system, for which 
 the  separation between two component is  $\sim0.1-0.2$~AU and the
 compact object is moving around an O6.5V main sequence star with a short  
orbital period $P_{ob}\sim 3.9$~days and a moderate eccentricity 
($e\sim0.24-0.35$, Casares et al. 2005; Aragona et al. 2009; Sarty et al. 2011). The binary system is a source of non-thermal emission 
in radio (Mold${\rm\acute{o}}$n et al. 2012), X-ray (Takahashi et al. 2009), 
and  gamma-ray (Abdo et al. 2009 for GeV; Aharonian et al. 2006 for TeV) 
bands, and exhibits  temporal variations in its emission and spectrum. 

The modulating GeV emission from LS~5039 has been confirmed by 
the $Fermi$-LAT (Abdo et al. 2009).  
The pattern of the orbital modulation of 
GeV emissions shows in anti-phase with X-ray and 
TeV gamma-ray emissions (c.f. Figure~\ref{light-ene}); 
the observed GeV flux  (or X/TeV fluxes) becomes maximum around the  superior 
conjunction (or inferior conjunction) and  becomes minimum around the 
inferior conjunction (or superior conjunction). The spectrum in 0.1-10GeV 
bands  is  harder when the emission is weaker.  
The phase-averaged spectrum shows 
a cut-off around $\sim 2$GeV, and but 
 Hadasch et al. (2012) found an emission feature above 10~GeV, which 
will be  compatible to lower energy tail of the TeV emissions.

The origin of the GeV emissions from the LS~5039 has been remained to be 
solved. Because the spectral shape of LS5039 measured by \fermi\ resembles to 
  those of the gamma-ray emitting pulsars, it has been suggested 
that LS~5039 includes a young pulsar and 
the emissions from the magnetosphere or  the cold-relativistic 
 pulsar wind produces the GeV emissions 
(Sierpowska-Bartosik \& Torres 2007; Kapala et al. 2010; 
Torres 2011). On the other hand, the inverse-Compton scattering process of 
 the pulsar wind accelerated 
by the inter-binary shock was also proposed to explain the GeV emissions 
(Yamaguchi \& Takahara 2012; Zabalza et al. 2013).

The main purposes of this study are (1)  to present results of the
 observational analysis using  4-year \fermi\ data, which provide us 
a more detailed information on the GeV emissions from LS~5039, 
and (2) to develop a model to discuss  
the emission processes of the X-ray, GeV and TeV gamma-rays. 
The present paper is composed of  two parts. 
In the first part, we report on results of four year observations of the 
\fermi.  Although Hadash et al. (2012) found the emissions above 10~GeV 
with 2.5yr $Fermi$ data, the  large uncertainty in the phase-resolved 
spectra in those energies prevents us to understand the detailed
 spectral behavior above 10~GeV. Furthermore, the emissions 
around 100MeV are strongly affected by the background model. 
In this paper, therefore, we perform a more detailed analysis with updated 
instrument response function to obtain 
more solid understanding of 
the spectral behavior in 100MeV and $>10$GeV energy bands.

In the second part, we will develop the emission model, in which 
the emissions from  magnetospheric outer gap and from cold-relativistic 
pulsar wind contribute to the GeV emissions of LS~5039, and will compare 
the predicted emission properties with the results of \fermi\ observation. 
 We will also study the X-ray and TeV gamma-ray  emissions 
from the intra-binary shock and will discuss the properties of the 
phase-resolved spectra in the multi-wavelength bands (X-ray, GeV and TeV).
 In section~\ref{observation}, we will report on the 
results of  our analysis of the four year \fermi\ data. We will 
 describe the emission model in section~\ref{model} and compare 
the model predictions with the results of the 
multi-wavelength  observations in section~\ref{result}. 
Discussion and a brief  summary are  given in  sections~\ref{discussion}
 and \ref{summary}, respectively.

\section{Data analysis and results of the \fermi\ data}
\label{observation}
\subsection{Data set}
In this study, we used data collected starting 2008 August 14 and extending 
until  2012 May 19. The observation time was limited by the availability of the
 timing model of the nearby gamma-ray pulsar, PSR J1826-1256, which was 
 needed for removing the contribution of the pulsar. The 
 timing model was adopted from  the \fermi\ LAT Multiwavelength Coordinating Group
 \footnote{https://confluence.slac.stanford.edu/display/GLAMCOG/LAT+Gamma-ray+Pulsar+Timing+Models} 
 \citep{ray11}. The data were reduced and analyzed using the \fermi\ Science
 Tools package (v9r32p5), available from the \fermi\ Science Support Center 
\footnote{http://fermi.gsfc.nasa.gov/ssc/data/analysis/software/}. We selected 
only events in the Reprocessed Pass 7 'Source' class and used the P7REP\_SOURCE\_V15 
version of the instrumental response functions. To reduce contamination from the 
Earth's albedo, we excluded time intervals when the region of interest (ROI) 
was observed at zenith 
angles greater than 100\degr\ or when the rocking angle of the LAT was greater
 than 52\degr. To minimize background from the nearby gamma-ray pulsar, 
PSR J1826-1256, we excluded events arriving in the pulse phase 
intervals 0.05-0.2 and 0.6-0.75 of the pulsar.

\subsection{Spectral analysis}
The {\it gtlike} tool was used for spectral analysis. We used photons between 0.1 and
 300 GeV within a $20\degr\times20\degr$ ROI centered at the position of 
LS 5039. For source modeling, all 2FGL catalog sources \citep{nolan12} within 
$19\degr$ of the ROI center, the galactic diffuse emission 
(gll\_iem\_v05.fit) and isotropic diffuse emission (iso\_source\_v05.txt)
 were included. 

\subsubsection{Phase-averaged spectrum}
\begin{figure}
\plotone{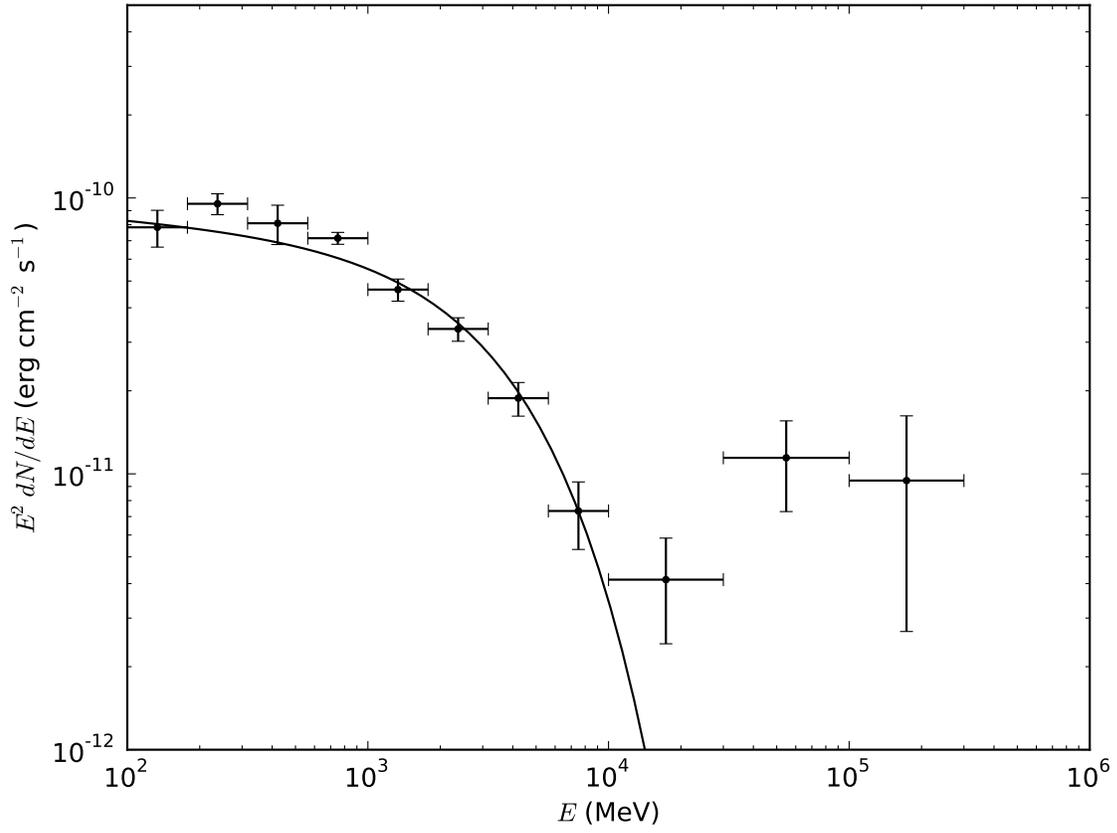}
\caption{Phase-averaged spectrum of LS 5039. The solid line shows 
the best-fit model for the full energy band.}
\label{spec}
\end{figure}
We modeled LS 5039 with a power law with an exponential cutoff
\begin{equation}
\frac{dN}{dE} = N_0 \left(\frac{E}{E_0}\right)^{-\Gamma} \exp\left(-\frac{E}{E_{cutoff}}\right).
\end{equation}
The spectral types of  other point sources in 2FGL are modeled
 according to the spectral types in the catalog, with the spectral parameters 
of sources more than $10\degr$ away from the ROI center fixed to the catalog 
values. The best-fit parameters are $\Gamma = 2.06 \pm 0.02_{stat}$ 
and $E_{cutoff} = 3.42 \pm 0.17_{stat}$ GeV.

Spectral points were obtained by performing a fit in each energy band, 
fixing the spectral parameters of sources more than 4\degr\ away from the ROI center, leaving the flux normalization constants of all other sources and the 
diffuse background free. The sources that were left free were modeled by 
power laws. In addition, an initial fit was performed and sources with 
TS $< 0$ in that energy band were removed. Figure.~\ref{spec} shows 
the phase-averaged spectrum. Significant emission is observed at E $>$ 10 GeV, in agreement with \citet{had12} and \citet{2fhl}.

\subsubsection{Phase-resolved spectra}
\label{pspec}

\begin{figure}
\plotone{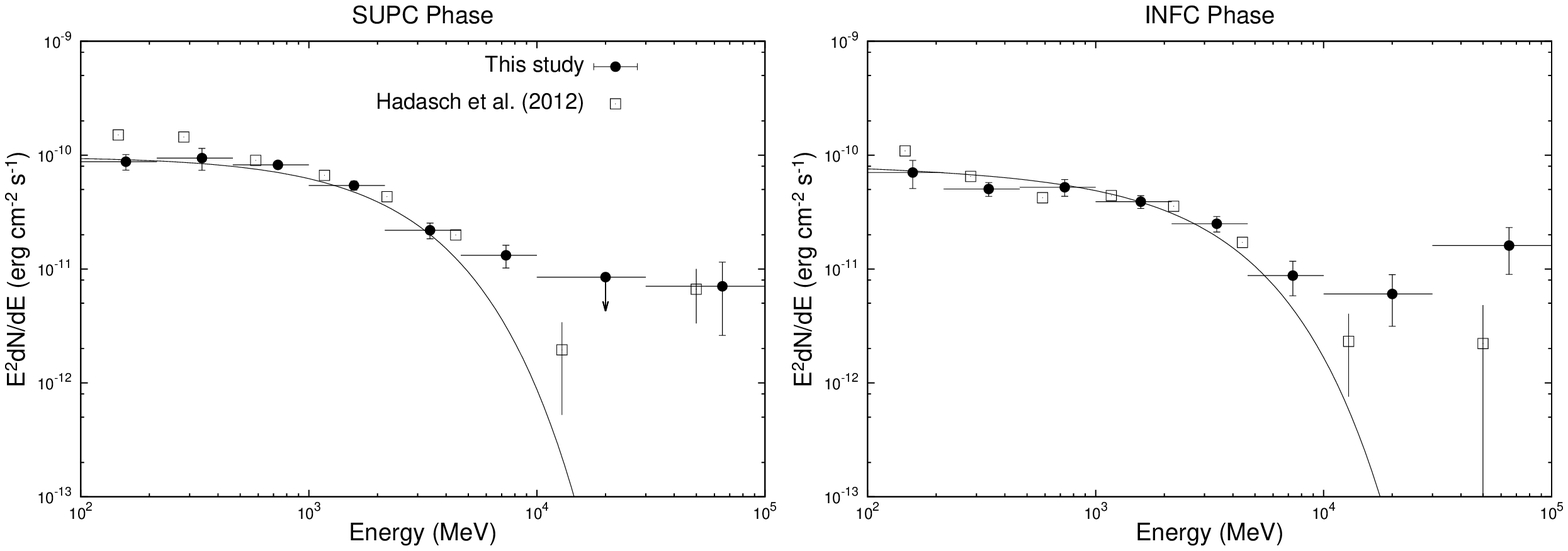}
\caption{Phase-resolved spectra of LS 5039 in SUPC phase (left) and 
INFC phase (right). The results obtained by Hadasch et al. (2012) are also 
displayed for comparison. The solid lines show the  bet-fitting functions 
described in section~\ref{pspec}.}
\label{spec2}
\end{figure}


We first performed phase-resolved analysis following the H.E.S.S. analysis 
by \citet{aha06}. We set phase zero at the 
periastron( $\phi=0$ with MJD=51942.59) and 
divided one orbit into two phases, that is,  the superior 
conjunction phase (SUPC phase, $0<\phi<0.45$ and $0.9<\phi<1$), which includes 
 SUPC ($\phi=0.06$),  and inferior 
conjunction phase (INFC phase, $0.45<\phi<0.9$), which 
includes INFC ($\phi=0.72$).  We performed similar likelihood analysis 
in the two phase intervals. In the fitting, all the spectral parameters of 
LS 5039 were left free, while other spectral parameters were fixed to the 
phase-averaged values, except for the flux normalization parameters of sources 
within 5\degr of the ROI center and the galactic and isotropic diffuse 
emissions. The best-fit parameters for the SUPC (INFC) 
are $\Gamma = 1.96 \pm 0.09_{stat}$ ($1.89 \pm 0.12_{stat}$) and 
$E_{cutoff} = 2.86 \pm 0.57_{stat}$ ($3.58 \pm 0.97_{stat}$) GeV. The spectrum is 
shown in Figure~\ref{spec2}. It is found that the flux in $\sim$100-300 MeV 
has noticeably decreased compared with previous studies done 
by Abdo et al. (2009) and  \citet{had12}. 
This can be due to the improvement in instrumental response functions 
 and data from Pass 6 to Reprocessed Pass 7.
 The flux beyond $\sim10$GeV is also 
increased compared with the previous studies, which is also seen in other 
results with the Reprocessed Pass 7 data (e.g. see \citealt{bregeon13}).  

\begin{figure}
\begin{center}
\includegraphics[height=8cm]{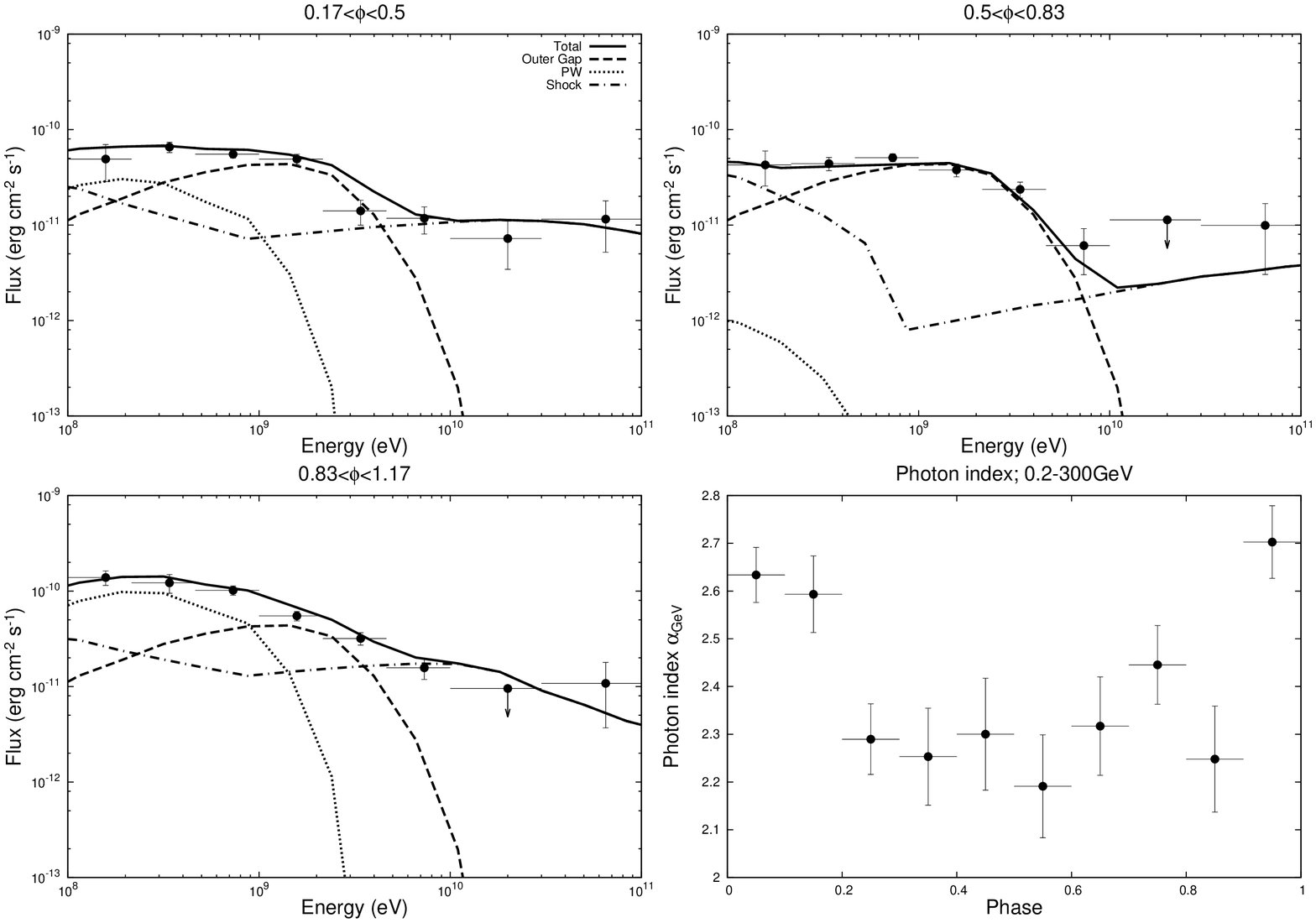}
\caption{Phase-resolved spectra and the photon index of GeV gamma-ray 
 emissions from LS~5039. 
Top-left, top-right and bottom-left show the spectra for the orbital 
interval $0.17<\phi<0.5$, $0.5<\phi<0.83$ and $0.83<\phi<0.17$, respectively.
Bottom-right shows the photon index fitted by a single power-law function 
on 0.2-300GeV.  The results of current \fermi\ data analysis are represented 
by the filled circles. The solid lines show the results of the theoretical 
model (section~\ref{model}), which includes emissions from the outer gap (dashed lines), 
the cold-relativistic 
pulsar wind (dotted lines) and the shocked pulsar wind (dashed-dotted lines). }
\label{fit}
\end{center}
\end{figure}

Due to the increased statistics in this study, we are allowed to divide
 the observation time into more orbital phase bins. We divided the observation 
time into three equally spaced orbital phase bins: $[0.17,0.5]$ (bin 1), 
$[0.5,0.83]$ (bin 2, INFC) and $[0.83,1]\cup[0,0.17]$ (bin 3, SUPC). Both the 
first and second intervals touch the apastron, but only the second interval 
contains the INFC.  This cut is chosen to better isolate the INFC from the 
apastron. In addition, the first and second phase bins exclude  
the SUPC and the third phase bin includes the emissions at the SUPC.
 The results are shown in Figure.~\ref{fit} for comparing 
with results of the theoretical model discussed in section~\ref{model}.
 We can see in the figure that the spectra of two orbital bins  excluding  
the superior conjunction (upper panels)  
have a clear spectral cut-off at $\sim 2$ GeV  and the spectra below 10GeV 
resemble each other. At the phase bin containing 
the SUPC (lower left panel), an enhancement at $0.1-0.3$ GeV 
is exclusively seen, and 
the spectrum below 10GeV is softer than other two phase bins. This suggests
 that the emissions below 10~GeV are composed of the two components, that is, 
one contributes to emissions around the SUPC in $<1$GeV bands and other 
dominates  in the  1-10GeV emissions for entire orbit.  
As we can see in Figure~\ref{fit}, the change of the spectral slope at 
 around 20~GeV in each phase bin suggests existence of an additional 
component that is compatible to lower energy tail of the TeV emissions

\begin{figure}
\begin{center}
\includegraphics[height=5cm]{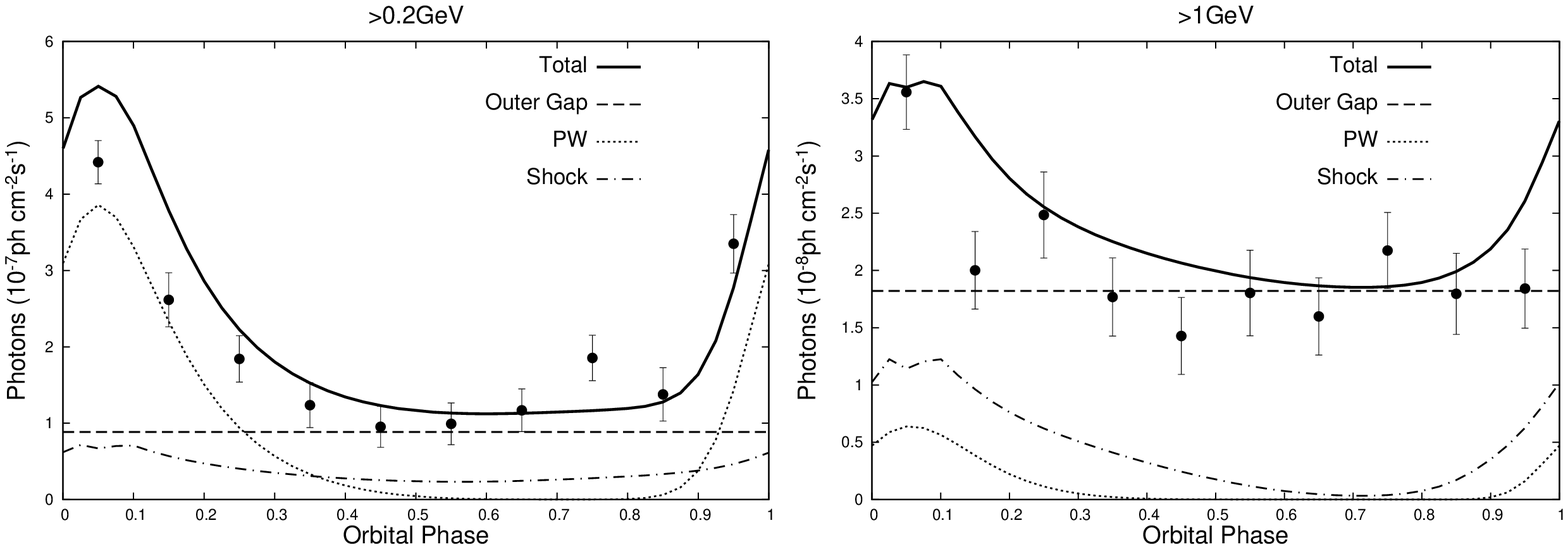}
\caption{Variation of the integrated flux in  $>0.2$GeV bands (left panel)
 and $>1$GeV bands (right panel). The symbols and the lines correspond to 
the same case as Figure~\ref{fit}. }
\label{light-fit}
\end{center}
\end{figure}
\subsection{Orbital light curves}

To obtain light curves, we performed likelihood analysis similar to 
Section \ref{pspec}. The orbital modulation for the flux 
 is summarized  in Figure~\ref{light-fit}, in which  the left  and 
right panels  display 
the light curves in two  energy bands, 0.2-300 GeV and 1-100 GeV, respectively.
In Figure~\ref{light-fit}, results of the theoretical model are also 
displayed for comparisons.  
The observed trends of flux modulation of the two energy bands 
are similar, but the amplitudes in the modulations are significantly different.
 The flux including lower energy photons  (0.2-300 GeV) 
is modulated by a factor of $\sim 5$ while 
that in the high energy band (1-100GeV) is modulated by only 
a factor of $\sim 2$. This indicates that the variation of spectral 
hardness along the orbital phase is 
mainly due to the modulation in the low energy band rather than the high 
energy band, and the high energy bands are dominated by a component which 
does not vary with the orbital phase.  This could also support the hypothesis 
that  the GeV emissions from LS~5039 are composed of several components
 with different characteristic energies.  
This feature is more clearly seen in our results than the results
 in previous studies. 
To qualitatively describe the evolution of the hardness of the emissions,
 we  extracted the photon index of each phase interval by 
 assuming a simple power law in 0.2-300 energy bands,
\begin{equation}
\frac{dN}{dE} = N_0 \left(\frac{E}{E_0}\right)^{-\Gamma}.
\end{equation}
 It is clear from Figure~\ref{spec} that there is 
a curvature in the spectra, and 
a single power law function would  not be  appropriate at some orbital-phases. 
However, because of the reduced statistics in individual phase bins, it is
 not possible to distinguish a simple power law from a power law with 
exponential cutoff with statistical significance in some bins. 
A fit with a simple power law can provide a quantitative
 indicator of the hardness of the spectrum. 
The result is shown in Figure.~\ref{fit} (lower right panel).
The spectrum is the softest around SUPC and  hardest in a 
broad region centered around the apastron ($\phi=0.5$).

In summary, the results of the four year observations by
$Fermi$ improve our understanding of the GeV emissions from LS 5039.
 The current results suggest that 0.1-100GeV emissions from LS~5039 
are likely composed of the
three different components (i) the first
 component contributes to $<$1GeV emissions around superior conjunction, 
(ii) the second component dominates in 1-10GeV energy bands for entire orbit 
and (iii) the third component  is compatible to lower energy
 tail of the TeV emissions.


\section{Theoretical model}
\label{model}
\begin{figure}
\begin{center}
\rotatebox{90}{\includegraphics[height=15cm,width=9cm]{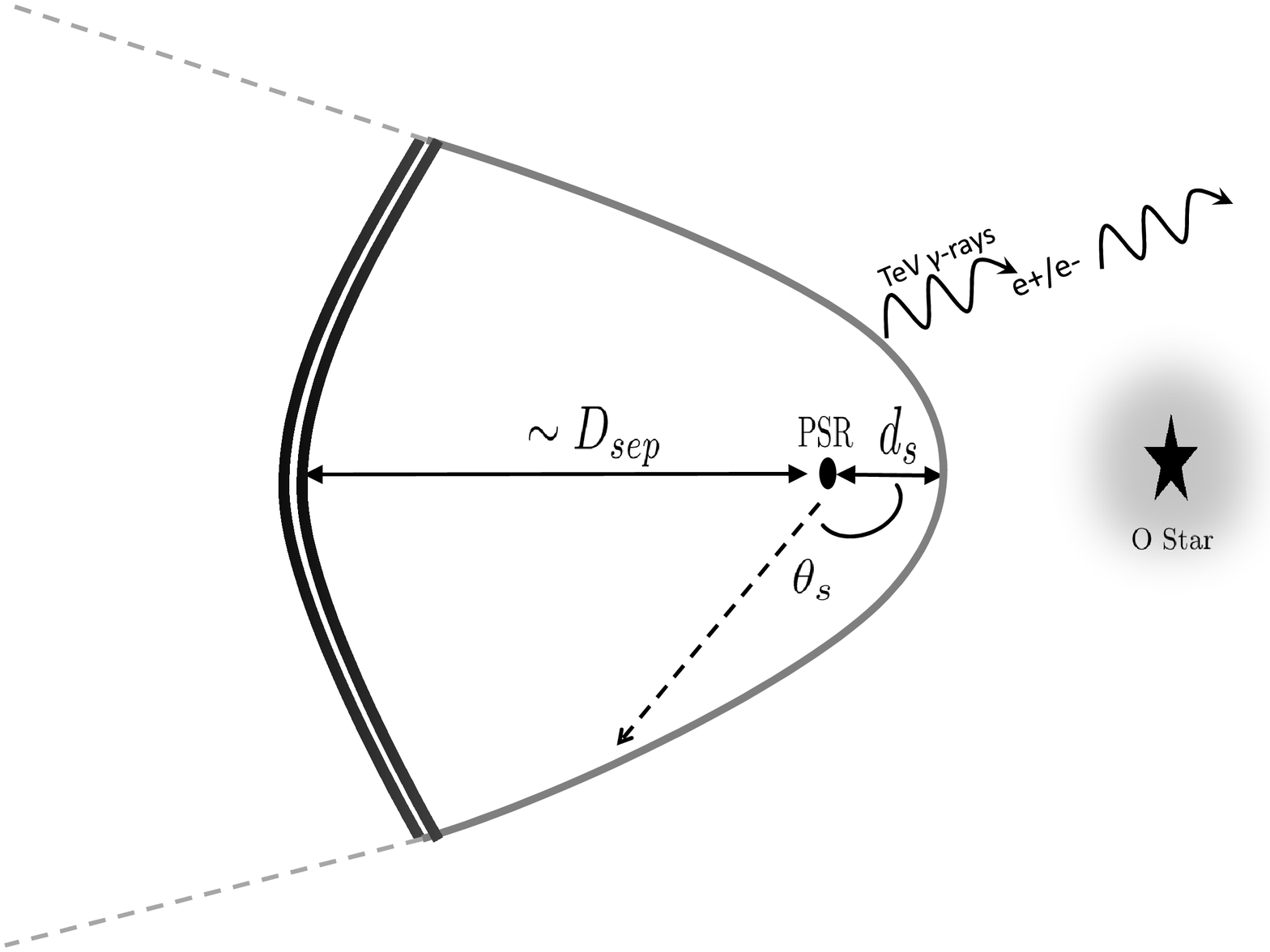}}
\caption{Schematic view of proposed picture of LS~5039 system. LS~5039 
comprises a pulsar and an O star, and  
the interaction between the pulsar wind and the stellar wind 
from the companion star forms a termination shock (Shock-I, thick-solid line). 
The distance to shock apex ($d_s$)  
from the pulsar is determined by the balance between the pulsar wind  pressure
 and stellar wind pressure. The effect of the orbital motion of the pulsar 
also produces a termination 
shock in opposite direction of the companion star (Shock-II, 
double-solid lines).  The distance to shock from the pulsar is of order 
of the orbital separation ($D_{sep}$), 
for which the  ram pressure of the stellar wind owing 
to Coriolis force and the ram pressure of the pulsar in balance.
The GeV emissions are mainly produced by the curvature radiation 
in the magnetospheric outer gap and by the inverse-Compton scattering 
of the cold-relativistic pulsar wind. The synchrotron process and 
inverse-Compton process of the shocked pulsar wind produce the X-rays 
and TeV gamma-rays, respectively.  
The TeV gamma-rays emitted from the shocked pulsar wind 
create new electrons and positron pairs, and they initiate the 
pair-creation cascade, if they propagate toward the companion star.}
\label{LS5039}
\end{center}
\end{figure}
 In the last section, we discussed the evidence of 
 two components of 
the emissions in  0.1-10GeV bands observed by the $Fermi$. In this paper, 
we will propose that the emissions in 1-10GeV energy band are 
 dominated by 
the magnetospheric emissions, while $\sim 0.1$GeV  gamma-rays 
 around the superior conjunction  are mainly produced by  the emissions 
from the cold-relativistic pulsar wind. Hence our model predicts 
that the emissions in 1-10 GeV energy band are pulsed. 
 
 Figure~\ref{LS5039} shows the schematic view of the 
LS~5039 system discussed in this paper.  Our 
 model assumes that LS~5039 includes a pulsar. The pulsar 
is wrapped by the termination shocks (section~\ref{shockgeo}),
 where the pulsar wind is stopped. The emissions from the 
magnetosphere (section~\ref{magemi})
 and the inverse-Compton scattering process of the 
cold-relativistic pulsar wind (section~\ref{cpw}) contribute to the observed 
GeV emissions. The synchrotron radiation and the inverse-Compton process
 of the shocked pulsar wind produce the X-rays and TeV 
gamma-rays, respectively (section~\ref{shocke}).

 We note that the shock geometry applied in this study 
(c.f. section~\ref{shockgeo}) resembles to 
one presented in  Zabalza et al. (2013), who applied two shock regions; 
the first shock (Shock-I) is located between the pulsar and the companion star,
 and the second shock (Shock-II)  is opposite direction from the companion star.
Both our model and Zabalza et al. (2013) assume  that TeV gamma-rays are mainly 
produced by the Shock-II region (c.f. section~\ref{multi}).
Main difference between us and Zabalza et al. (2013) is  X-ray/GeV emission 
processes. Our model will predict that the particles 
accelerated at the Shock-I  produce the X-rays via 
synchrotron radiation, while  Zabalza et al. (2013) proposed that 
the  particles accelerated at Shock-I  
produce the Gev emissions via the IC process.   
For  emissions in 0.1-10GeV bands, we expect that the emissions are 
mainly produced by the magnetospheric particles in the outer gap and 
the cold-relativistic pulsar wind. Since Zabalza et al. (2013) mainly focused 
 on the GeV/TeV gamma-ray emission process and had a difficulty to explain 
the X-ray emission process,  we will 
develop an emission model covering from X-ray to TeV energy  bands.

\subsection{Shock geometry}
\label{shockgeo}
We assume that two kinds of termination 
shock exist around the pulsar (c.f. Figure~\ref{LS5039}). 
First,  balancing between the 
pressures of the pulsar wind and the stellar wind produces cone-shape shock 
 between the pulsar and the companion star (Shock-I). The opening angle 
and the geometry 
of the cone-shape shock are determined by ratio of the momenta of the 
two winds (e.g. Eichler \& Usov 1993; Canto et al. 1996),
\begin{equation}
\eta\equiv \frac{L_{sd}}{\dot{M}v_w c},
\end{equation}
where $L_{sd}$ is the spin down energy, 
$\dot{M}$ is the mass loss rate of the outflow 
from the companion star and $v_w$ is the velocity of the outflow.  
The distance to the shock apex from the pulsar can be determined by 
$r_{apex}=D_{sep}\eta^{1/2}/(1+\eta^{1/2})$, where $D_{sep}$ is the separation 
between two stars. In this paper, we will apply $\eta=0.05$ with
 $L_{sd}\sim 2\times 10^{36}{\rm erg~s^{-1}}$,
 $\dot{M}\sim 10^{-7}M_{\odot}{\rm yr^{-1}}$ and 
 $v_w\sim 2\times 10^8 {\rm cm~s^{-1}}$.

\begin{figure}[h]
\begin{center}
\includegraphics[height=9cm,width=9cm]{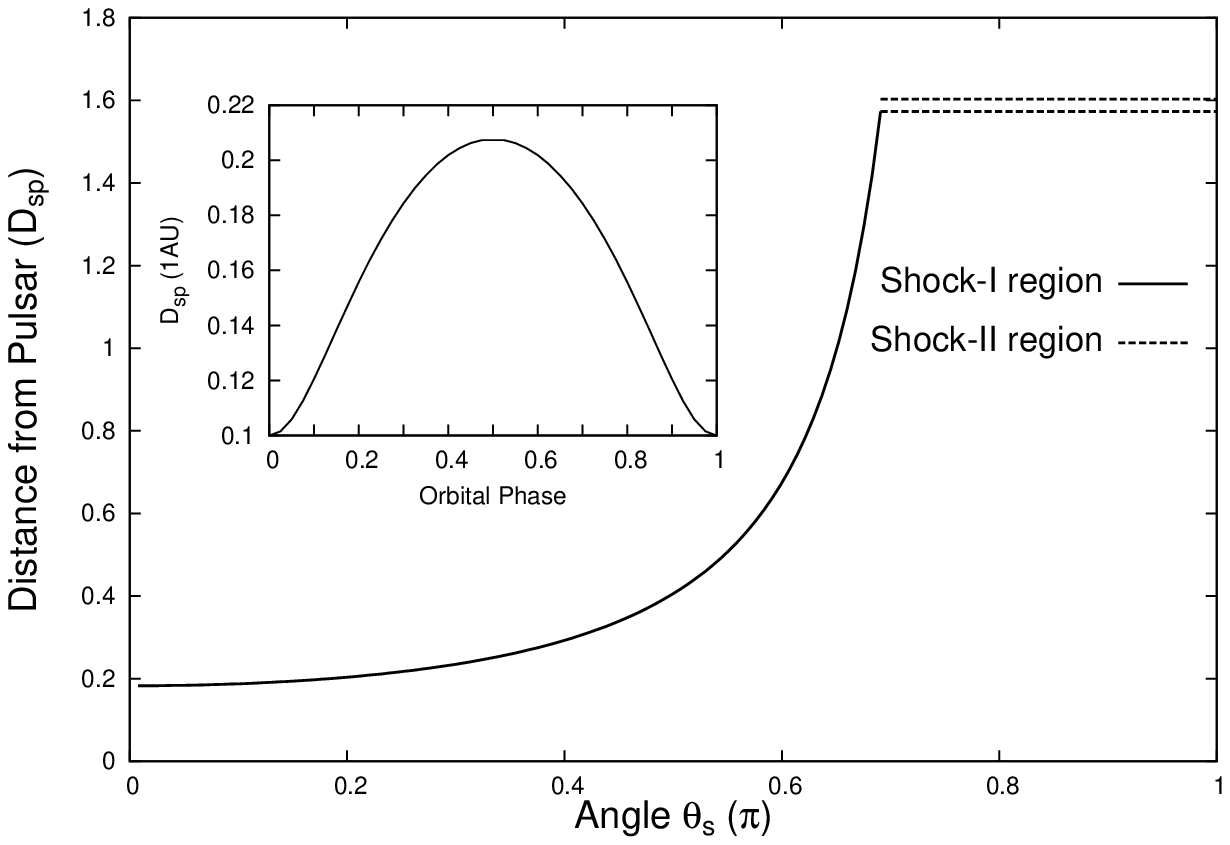}
\caption{Distance to termination shock from the pulsar with respect to 
the angle measured ($\theta_{s}$, c.f. Figure~\ref{LS5039})
 from the direction of the companion star. Solid line: Distance to the 
shock caused by the interaction between 
the pulsar wind and the stellar wind (Shock-I). 
The result is for the momentum ration of 
$\eta=0.05$. Double solid line: Distance to the shock caused by 
the effect of the orbital motion of the pulsar (Shock-II). In the calculation, 
we assume $r_s=1.5D_{sep}$. The sub-panel in the figure show the separation 
between two stars as a function of the orbital phase.}
\label{line}
\end{center}
\end{figure}

Second, recent results of 2-D hydrodynamic simulation
 of LS~5039 (e.g. Bosch-Ramon et al. 2012) have suggested that the effect 
of the orbital motion also produces  a pulsar wind 
termination shock even in opposite direction from the companion star 
 (Shock-II, c.f. double-solid line in Figure~\ref{LS5039}). 
The distance from the pulsar is found to be of order of the orbital 
separation, for which the  ram pressure of the stellar wind owing 
to Coriolis force and the ram pressure of the pulsar are in balance.  
 Zabalza et al. (2013) provided an approximate expression for 
location of the shock as 
\begin{equation}
x_{cor}\sim \sqrt{\frac{L_{sd}v_w}{\dot{M}c(2\Omega_o)^2}},
\label{xco}
\end{equation}
where $\Omega_o$ is the angular velocity of the pulsar around the star. 
With $\Omega_o\sim 10^{-5}~{\rm s^{-1}}$, the expression~(\ref{xco}) yields 
$x_{cor}\sim 0.15{\rm AU}\sim 1-2D_{sep}$. In this paper, 
we assume $r_s= 1.5D_{sep}$ as 
the radial distance to  the Shock-II from the pulsar. 
 Figure~\ref{line} shows  the distance to the shock as a function
 of angle  measured from the direction of the companion star. 

 It has been pointed out that 3D simulation 
will give a more complex  structure of Shock-II, because the instability 
will develop faster and more disruptive in 3D simulation 
than 2D (Bosch-Ramon et al. 2012). Because the detailed structure of 
the shock-II region given by the 3D simulation has not been known 
for LS~5039 system, we apply the result of 2-D calculation in this study.

\subsection{Magnetospheric emission}
\label{magemi}
We calculate the expected spectra of the gamma-ray emissions from 
 the outer gap accelerator in the pulsar magnetosphere  (Cheng et al. 1986). 
In the outer gap, the electrons and positrons are 
accelerated by the electric field along the magnetic field lines. 
The accelerated particles emit the GeV gamma-rays through the curvature 
radiation process.  The typical magnitude of the accelerating electric field is  
\begin{equation}
E_{||}\sim \frac{f_{gap}^2 B(R_{lc})R_{lc}}{R_c},
\end{equation}  
where $B$ is the local magnetic field strength, 
$R_{lc}=c/\Omega$ is the radius of the light cylinder, $R_c$ is curvature 
radius of the magnetic field line, and
  $f_{gap}$ is the ratio of the gap thickness in the trans-field 
direction and the light cylinder radius at the  light cylinder. 
 The accelerated electrons and positrons  
produce gamma-rays via the curvature radiation process. The Lorentz 
factor of the electrons/positrons is estimated by balancing between 
the acceleration force and back reaction force of the curvature radiation, 
\begin{equation}
\Gamma_e\sim \left(\frac{3R_c^2}{2e}E_{||}\right)^{1/4}\sim 
2\times 10^7\left(\frac{f_{gap}}{0.1}\right)^{1/2}
\left(\frac{\Omega}{10^2{\rm s^{-1}}}\right)^{1/2}
\left(\frac{B(R_{lc})}{10^5{\rm G}}\right)^{1/4}
\left(\frac{R_{c}}{R_{lc}}\right)^{1/4}.
\end{equation}
Typical energy of the curvature radiation is found to be 
 of order of $\sim$GeV,  
\begin{equation}
E_c\sim \frac{3hc\Gamma_e^3}{4\pi R_c}
\sim 800 \left(\frac{\Gamma_e}{2\cdot 10^7}\right)^3
\left(\frac{\Omega}{10^2{\rm s^{-1}}}\right)
\left(\frac{R_{c}}{R_{lc}}\right)^{-1} {\rm MeV}.
\end{equation}
The luminosity of the gamma-ray emissions from the outer gap 
becomes 
\begin{equation}
L{\gamma}\sim f^3_{gap}L_{sd}\sim 3.8\times 10^{35}f_{gap}^3
\left(\frac{P}{0.1{\rm s}}\right)^{-4}
\left(\frac{B_s}{10^{12}{\rm G}}\right)^{2} {\rm erg~s^{-1}},
\label{lgamma}
\end{equation} 
where $L_{sd}=2(2\pi)^4 B^2_s R^6_s/(3c^3P^4)$ is the spin down power,
 $P$ is the pulsar's spin period, $B_s$ is the stellar magnetic 
field and $R_s=10^6$cm is the radius of neutron star. The fractional gap 
thickness is estimated as
\begin{equation}
f_{gap}={\rm min}(f_m, f_p),
\label{fgap}
\end{equation}
 where $f_m=0.25K(P/0.1{\rm s})^{1/2}$ with $K\sim 2$ and 
$f_{p}=5.5P^{26/21}(B_s/10^{12}{\rm G})^{-4/7}$ (Takata et al. 2010). A 
more detailed description of the outer gap model can be found  
in  Wang et al. (2010).
\subsection{Inverse-Compton emission from cold-relativistic pulsar wind}
\label{cpw}
It is possible that the inverse-Compton (IC)  
scattering  of cold-relativistic pulsar wind 
 off the stellar photons produces  high-energy gamma-rays 
(Ball \& Kirk 2000;  Khangulyan et al. 2011, 2012). 
For LS~5039 system,  this IC component could play an important 
role to explain for the observed emissions.  As we will show in 
equation~(\ref{gamma0}), the typical Lorentz factor of the cold-relativistic 
pulsar wind will be $\Gamma_0\sim 10^{4}$. Hence, the inverse-Compton 
scattering process will be occurred in the Thomson regime.  
 With  typical separation of two stars ($D_{sep}\sim 0.1$~AU),  
 the optical depth of IC process will be of order of unity (by ignoring 
the effect of the collision angle) 
\begin{equation} 
\tau_{IC}(D_{sep}=0.1{\rm AU})\sim n_{ph}\sigma_{IC}r_{s}            
\sim 5\left(\frac{\sigma_{IC}}{\sigma_{T}}\right) 
\left(\frac{T_{eff}}{3\cdot 10^4{\rm K}}\right)^3\left(\frac{R_{*}}{9R_{\odot}} 
\right)^2\left(\frac{D_{sep}}{0.1{\rm AU}}\right)^{-2}
\left(\frac{r_s}{0.01{\rm AU}}\right),                                
\end{equation}                    
where $r_s$ is the distance to the shock, $T_{eff}$ is the effective temperature 
of the companion, and $\sigma_T$ is the
 Thomson cross section. The luminosity becomes of order of 
\begin{equation}
L_{IC}\sim \frac{\Gamma_0kT_{eff}}{m_ec^2}L_{sd}
\sim 0.03\left(\frac{kT_{eff}}{3{\rm eV}}\right)\left(\frac{\Gamma_0}{5\cdot 10^3}\right)L_{sd}.
\end{equation}
Hence  the IC process of the  cold-relativistic will 
 produce observable  high-energy gamma-rays.

The characteristic energy of the IC photons depends on 
the Lorentz factor 
of the cold-relativistic pulsar wind.  
If the pulsar wind is a kinetically dominated flow, 
 the  typical Lorentz factor of the bulk flow will be 
\begin{equation}
\Gamma_{W,0}\sim \Gamma_{W,L}\sigma_L\sim 10^4\left(\frac{\kappa}{10^5}\right)^{-1}\left(\frac{L_{sd}}{10^{36}{\rm erg~s^{-1}}}\right)^{1/2},
\label{gamma0}
\end{equation}
where $\Gamma_{W,L}$  and $\sigma_L=B^2(R_{lc})/[4\pi \Gamma_{W,L}\kappa m_ec^2 n_{GJ}(R_{lc})]$ are the Lorentz factor and the magnetization parameter of the pulsar wind at 
the light cylinder, respectively. In addition, $n_{GJ}(R_{lc})=\Omega 
B(R_{lc})/(2\pi ce)$ is the Goldreich-Julian number density and $\kappa$ is 
the multiplicity. The observed power of the synchrotron nebulae around 
the pulsars implies  a multiplicity  of $\kappa=10^{4-5}$ 
(De  Jager et al. 1996; De Jager 2007; Harding \& Muslimov 2011).  
The radiation per unit energy power unit solid angle of single particle with 
a Lorentz factor $\Gamma_{W}$ is given by 
 \begin{equation}
\frac{dP_{IC}}{d\Omega_1}={\cal D}^2\int_0^{\theta_c}(1-\beta\cos \theta_0)I_b/h
\frac{d\sigma'}{d\Omega'}d\Omega_0,
\end{equation}
where $d\sigma'/d\Omega'$ is the differential Klein-Nishina cross section, 
${\cal D}=\Gamma_W^{-1}(1-\beta\cos\theta_1)^{-1}$,  $\theta_1$ and $\theta_0$ 
describe  the angle between the direction of the particle motion 
and the propagating direction of the scattered photons and background photons, 
respectively. In addition, $I_b$ is the stellar photon field and 
$\theta_c=\sin^{-1}R_*/r$ expresses the angular size of the star as seen 
from the point $r$. For the target  stellar photon field, 
we take the stellar radius $R_*=9R_{\odot}$, where  $R_{\odot}$ is  the 
solar radius, and an effective temperature $kT_{eff}=3.4$eV. 
With $kT_{eff}=3.4$ eV, the scattering process of the electrons with 
a Lorentz factor of $\gamma_e\ge 2\times 10^{5}$ is occurred in 
the Klein-Nishina regime (c.f. Figure~\ref{radloss}), implying 
existence of a break in the inverse-Compton spectrum 
 at around $\sim 10^{11}$eV.

A mono-energetic assumption for the distribution of electrons and positions 
in the pulsar wind had been assumed as a first approach to the problem
 (e.g.  Takata et al. 2009). It is suggested however that the energy 
distribution as a result of a dissipation of the magnetic energy 
to the particle energy can be different from the mono-energetic 
distribution (Sierpowska-Bartosik \& Torres 2008 and reference their in). 
In this paper, we explore the emissions with a relativistic Maxwell 
distribution of the form,
\begin{equation}
f(r, \Gamma_W)=K(r)\Gamma^2_W {\rm exp}\left(-\frac{3\Gamma_W}{\Gamma_0(r)}
\right), 
\label{dist}
\end{equation}
which provides the averaged Lorentz factor of 
\[
<\Gamma_W(r)>\equiv \frac{\int_1^{\infty}\Gamma_W f(r,\Gamma_{W})d\Gamma_W}
{\int_1^{\infty} f(r,\Gamma_{W})d\Gamma_W}\sim \Gamma_0 (r).
\]

We assume that the distance ($r_i$) from the pulsar at which 
the kinetically dominated pulsar wind is formed is smaller than 
the shock distance, and that the averaged Lorentz
 factor at $r=r_i$ is $\Gamma_0(r_i)=5\times 10^3$.
 The normalization $K(r_i)$ at $r=r_i$
 is calculated from
\[
m_ec^2\int_1^{\infty}\Gamma_Wf(r_i,\Gamma_W)d\Gamma_W= \frac{L_{sd}}{4\pi r_i^2 c}.
\]

\begin{figure}
\begin{center}
\includegraphics[height=9cm,width=9cm]{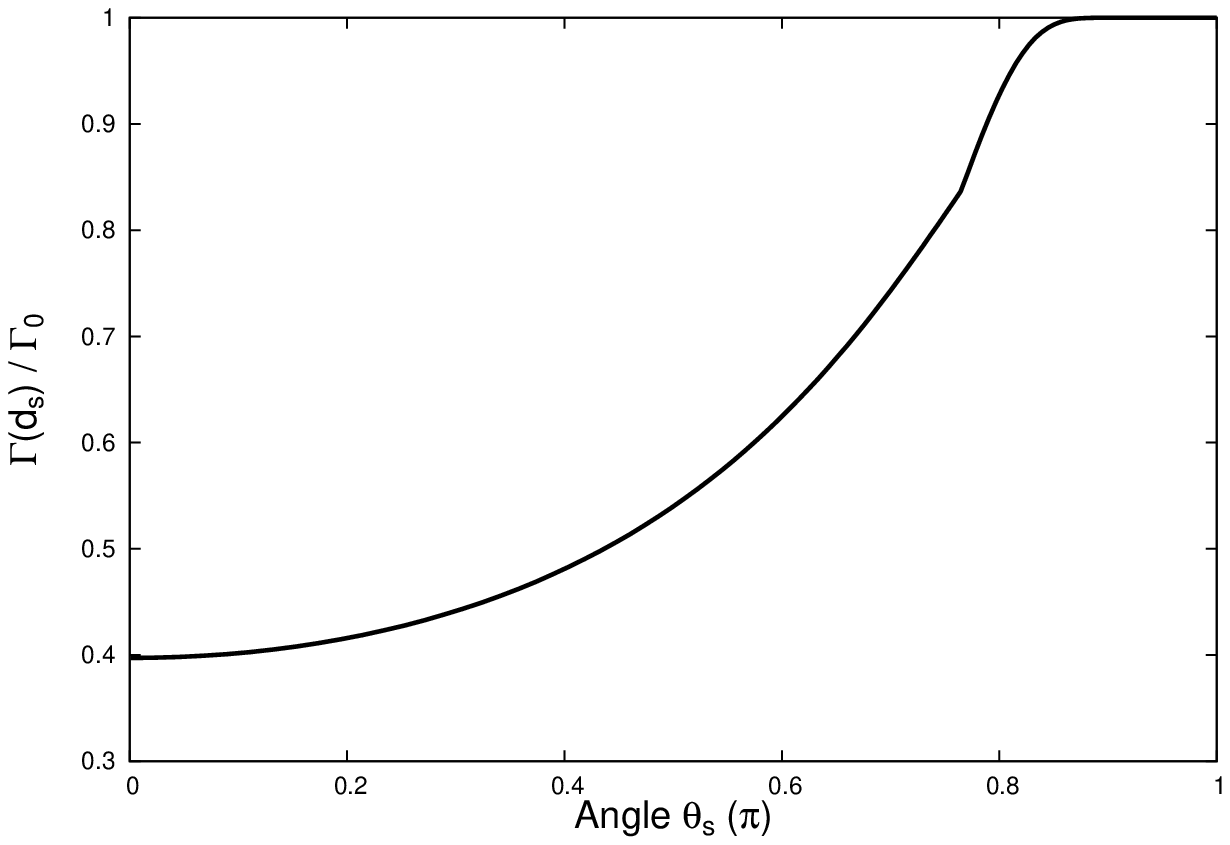}
\caption{The ratio of Lorentz factor at the shock 
 and initial Lorentz factor of the cold-relativistic pulsar 
wind with respect to the angle measured from 
the direction of the companion star. The cold-relativistic pulsar wind looses 
the  energy via the inverse-Compton scattering off the stellar photons.}
\label{gamma}
\end{center}
\end{figure}

The Lorentz factor 
of the pulsar wind evolves with the distance due to the energy loss by 
IC scattering process.  For example, Figure~\ref{gamma} shows the Lorentz factor
 of the pulsar wind at the shock as a function of the 
 angle measured from the direction of the companion star, where 
the initial Lorentz factor is $\Gamma_0=5\times 10^3$.  
We can see in Figure~\ref{gamma} that for the cold-relativistic 
pulsar wind propagating toward the companion star ($\theta_s\sim 0$),  
 ~40-50\% of the initial energy is released  before 
the shock, while  for the pulsar wind propagating in opposite direction  
of the companion star ($\theta_s\sim \pi$), 
the energy loss is negligible.  

We assume that at each point, ``thermalization'' is quickly established 
and the distribution is described by the relativistic Maxwell 
distribution~(\ref{dist}). The radiation power integrated  within 
the distance $r$ is 
\[
\delta P_{IC}=4\pi\int\int\int r^2f(r,\Gamma_W)
\frac{dP_{IC}}{d\Omega_1}d\Omega_1dr
d\Gamma_WdE_{\gamma},
\]
where $E_{\gamma}$ is the energy of scattered photons. The normalization 
$K(r)$ and the averaged Lorentz factor $\Gamma_0(r)$ are calculated 
with the equations of particle conservation and of the energy conservation, 
that is, 
\begin{equation}
r^2\int f(r,\Gamma)d\Gamma=r_i^2\int f(r_i,\Gamma)d\Gamma ,
\end{equation}
and 
\begin{equation}
m_ec^2\int \Gamma_Wf(r,\Gamma_W)d\Gamma_W+
\frac{\delta P_{IC}}{4\pi r^2c}= \frac{L_{sd}}{4\pi r_i^2 c},
\end{equation}
respectively.

We would like to mention that we additionally  calculated the emissions 
with mono-energetic distribution and  single power law distribution of
 the pulsar wind as well.
 For the mono-energetic distribution, we found that the shape of 
the calculated spectra resembles to one calculated with the relativistic
 Maxwell function. For the power law distribution, we assumed that 
the particles are accelerated above $\Gamma_{0} (r_i)\sim 5\times 10^3$ with 
a power law index of $3-4$, which has been 
predicted by the plasma simulations of the magnetic reconnection process
 (Zenitani \& Hoshino 2005).  We found that although the inverse-Compton 
spectrum extends to very high energy bands, its contribution with 
the ``soft''  power law index $3-4$  is much smaller than the shock 
emissions. Within the present framework of the particle distributions, 
therefore, the main results discussed in section~\ref{result} are not 
modified.

\begin{figure}
\begin{center}
\includegraphics[height=7cm,width=10cm]{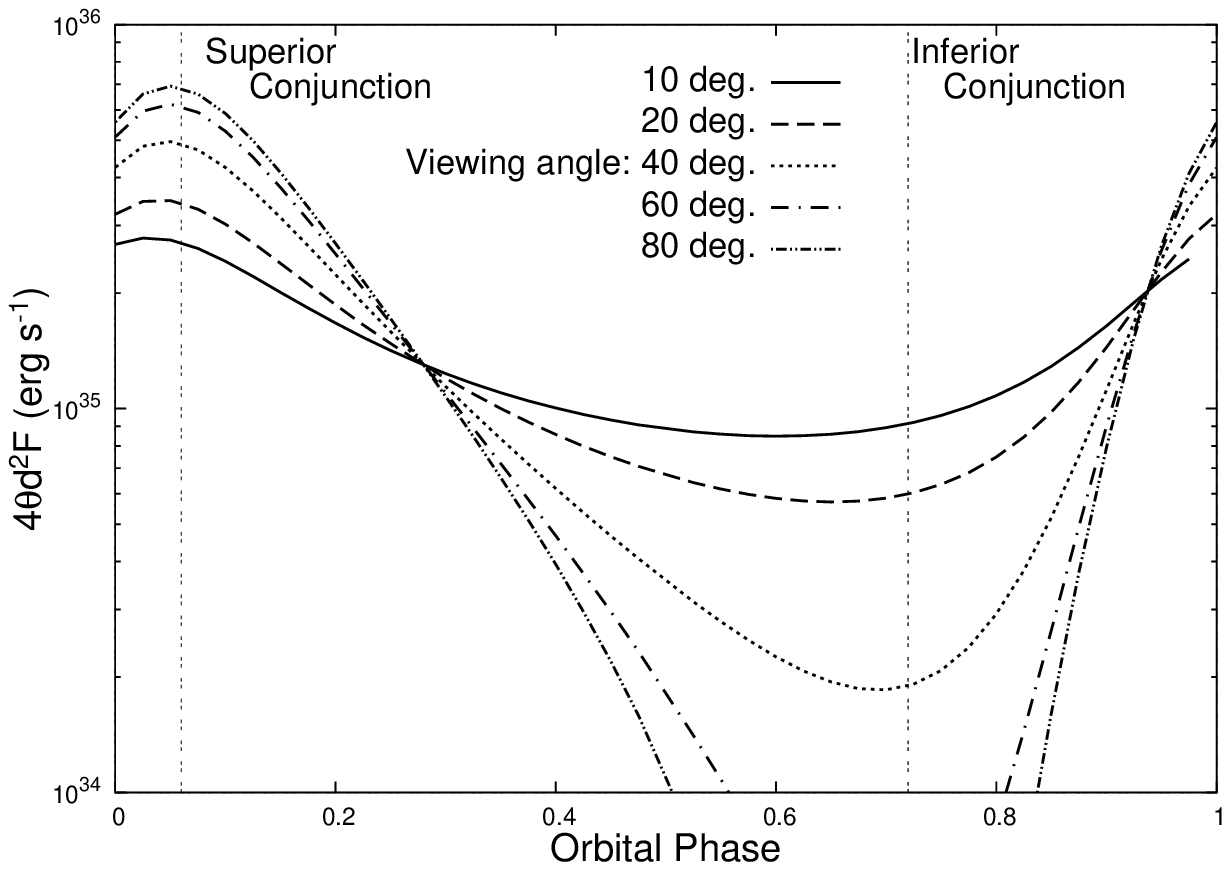}
\caption{Variation of the integrated flux of the inverse-Compton photons 
from the cold-relativistic pulsar wind. The results are for typical 
Lorentz factor of   $\Gamma_{W,0}=5\times 10^3$. 
The different lines show the results 
for the different Earth viewing angles measured from the direction
 perpendicular to the orbital plane. The phase zero $\phi=0$ corresponds to 
periastron, and the vertical double-dotted lies indicate 
the positions of the superior conjunction ($\phi=0.06$) and 
the inferior conjunction ($\phi=0.72$), respectively.}
\label{ic}
\end{center}
\end{figure}

Figure~\ref{ic} summarizes 
the temporal variations of the integrated flux of IC emissions  with respect  
 to the orbital phase. The different curves represent the results 
for different   Earth viewing angles measured from
 the direction perpendicular to the orbital plane.   
As we can see in Figure~\ref{ic}, the model light curves tend to have 
 a peak around  SUPC. This is (1) because the IC photons emitted toward 
the Earth are produced by the head-on like collision process,  and 
 (2) because  SUPC ($\phi\sim 0.06$) is close to the periastron ($\phi=0$), 
where the separation between two stars becomes minimum and hence 
the soft photon number density at the location of the pulsar 
becomes maximum. As a result, the IC process is  more efficient 
around SUPC.  Around INFC, the flux becomes minimum, since the 
 IC photons traveling  toward the Earth are produced by the tail-on 
collision process. In Figure~\ref{ic}, 
we can  see that a larger Earth viewing angle predicts 
a larger amplitude of  the modulation. This is because as the Earth 
viewing  angle approaches to the edge on, the amplitude of 
variation of the collision angle, which is angle  between the stellar photons
 and the pulsar wind that emits photons toward the Earth,
 along the orbital phase increases.  We also find a tendency 
in Figure~\ref{ic} that as Earth viewing angle becomes small, 
the positions of flux maximum and minimum shift 
toward the periastron ($\phi=0$) and apastron ($\phi=0.5$), respectively. 
This is related to the fact that  IC emissivity depends
 on (1) the collision angle and (2) the number density of the soft photons.
  For a larger Earth viewing angle, the effect 
of the variation of the collision angle with  the orbital phase
 affects  more to the variation of the  IC flux. 
In such a case, the flux maximum (or minimum) appears 
at the superior conjunction (or inferior conjunction). 
If the Earth viewing angle approaches to  zero,
the orbital variation of the collision angle is small, 
 and the variation of  number density of soft photons at the emission regions 
 mainly causes  the variation of  the IC flux. In such a case, 
the IC flux becomes  maximum (or minimum) 
 at the periastron (or apastron), where  
the photon number density at the emission region becomes  maximum (or minimum).

\subsection{Shock Emissions}
\label{shocke}

At the shock, the kinetic energy of the pulsar wind is converted into
the internal energy of the wind, and the distribution of particles
at the shock is assumed to be described by a power law over several
decades in energy. The minimum Lorentz factor ($\gamma_{e,min}$) 
of the shocked pulsar wind particles is assumed to be the average Lorentz 
factor of  the cold-relativistic pulsar wind at the shock 
($\Gamma_0(d_s)$, c.f. Figure~\ref{gamma}).  
The maximum Lorentz factor $\gamma_{e,max}$ is 
determined by  balancing between the acceleration time scale 
$t_a\sim \gamma_e m_ec/(eB)$ and the synchrotron loss time scale 
$t_s\sim 9m_e^3c^5/(4e^4B^2\gamma_e)$. In this paper, we will assume $p=2.1$ for 
 the power law index of the distribution of the particles accelerated at 
the shock. 

Since the ratio of the  pulsar wind momentum to the stellar wind momentum 
is  much smaller than unity ($\eta\ll 1$),  we approximate that the flow 
of the shocked pulsar wind points radially outwards 
from the companion star. In this study, we assume  
that velocity ($v_{PW}$) of the bulk motion of  shocked pulsar wind 
does not change with the radial distance, that is, $v_{PW}(r)=$constant, 
 because the high-energy emission occurs 
in the vicinity of the shock. 

 In down stream region, the particles loose their energy via 
 the cooling processes.  With the steady state approximation, 
the evolution of the distribution function $N(r,\gamma_e)$  is given by
\begin{equation}
\frac{\partial N}{\partial r}+\frac{\partial}{\partial \gamma_e}
\left(\frac{d\gamma_e }{dr}N\right)=Q(\gamma_e)\delta(r-r_s),
\end{equation}
where $Q(\gamma_e)$ is the source function at the shock. The energy 
loss of the particles is  calculated from
\begin{equation}
\frac{d\gamma_e}{dr}=\frac{1}{v_{PW}}\left[\left(\frac{d\gamma_e}{dt}\right)_{ad}
+\left(\frac{d\gamma_e}{dt}\right)_{syn}+\left(\frac{d\gamma_e}{dt}\right)_{IC}
\right].
\end{equation} 
We apply the adiabatic loss given by
\begin{equation}
\left(\frac{d\gamma_e}{dt}\right)_{ad}=\frac{\gamma_e}{3n}\frac{dn}{dt}
=-\frac{2v_{PW}}{3r},
\end{equation}
 where $n$ is the particle number density and we apply 
$nr^2=$constant. The synchrotron loss is given as 
\begin{equation}
\left(\frac{d\gamma_e}{dt}\right)_{syn}=-\frac{4e^4B^2\gamma_e^2}{9m_e^3c^5},  
\end{equation}
where we used the average pitch angle, because we expected that 
the magnetic field in
 the shocked pulsar wind is easily randomized. The IC energy loss rate is 
\begin{equation}
\left(\frac{d\gamma_e}{dt}\right)_{IC}=-\int\int(E-E_S)
\frac{\sigma_{IC}c}{m_ec^2E_s}\frac{dN_s}{dE_s}dE_sdE,
\end{equation}
where $dN_s/dE_s$ is the stellar photon field distribution and $\sigma_{IC}$ is 
the cross section for the isotropic photon field.  
We estimate the magnetic field just behind the shock as   
\begin{equation}
B(r_s)=3\left(\frac{L_{sd}\sigma(r_s)}{r_s^2c[1+\sigma(r_s)]}\right)^{1/2},
\end{equation}
where $\sigma(r_s)$ is the magnetization parameter at the shock. 
In down stream region, we consider that the magnetic 
field evolves as $B(r)r=$constant. 

We assume that the magnitude of 
the magnetization parameter at the shock depends on the shock distance from 
the pulsar, because  we expect that an energy conversion 
from the magnetic field to the particle energy of the cold-relativistic pulsar wind gradually decreases the magnetization parameter with 
the radial distance from the pulsar.  Because there is a theoretical 
uncertainty for the 
evolution of $\sigma$ with the radial distance,  
we describe the magnetization parameter at the shock with 
 a single power-law function, 
\begin{equation}
\sigma (r_s)=\sigma(r_{apex,0}) \left(\frac{r_s}{r_{apex,0}}\right)^{-\alpha},
\end{equation} 
where $r_{apex,0}$ is the shock apex distance at the periastron. The evolution 
of the magnetization parameter for the gamma-ray binary
 PSR B1259-63/LS 2883 system is also suggested to explain 
the X-ray/TeV emissions  (Takata \& Taam 2009; Kong et al. 2011, 2012). We will 
argue in section~\ref{depmag} that  the index $\alpha$ affects the predicted 
flux  at TeV energy band. 

\begin{figure}[h]
\begin{center}
\includegraphics[height=7cm,width=7cm]{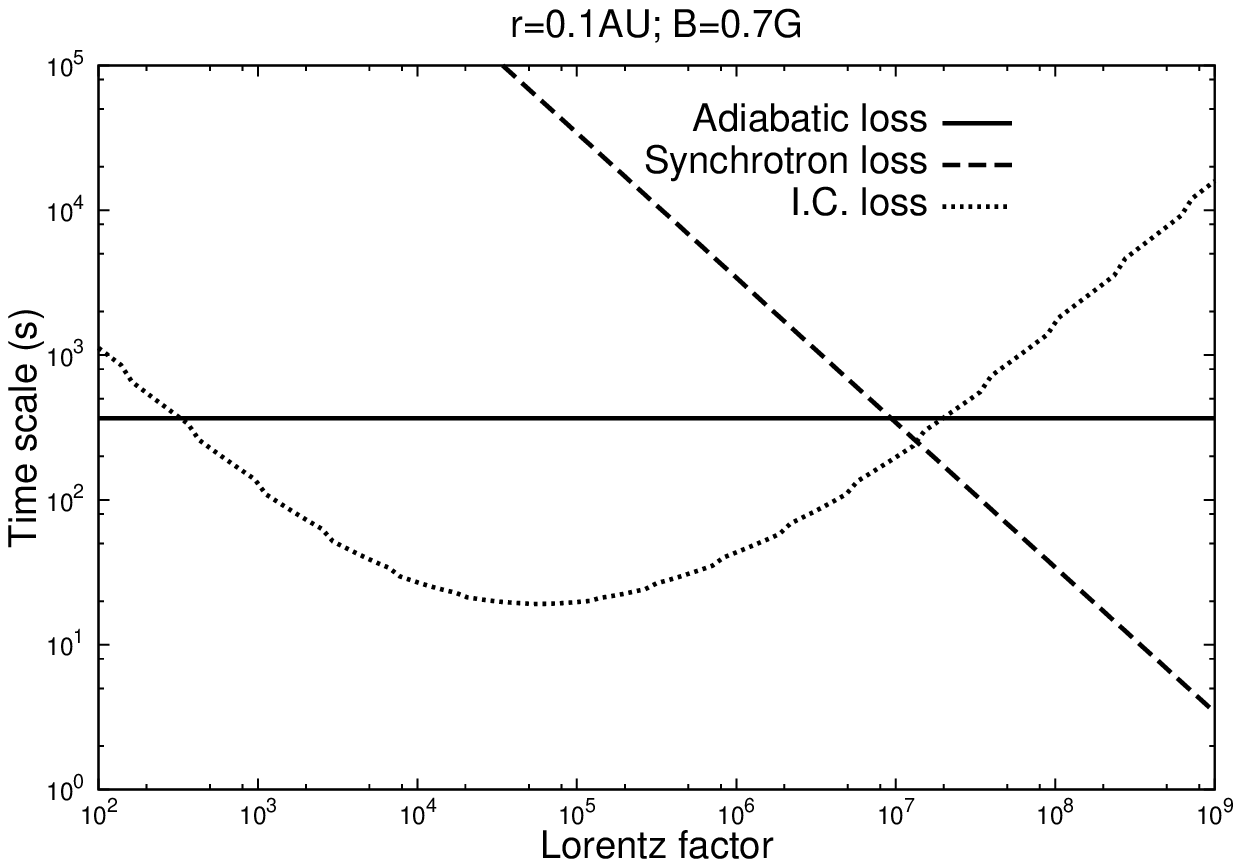}
\caption{Time scales of the adiabatic loss (solid line), 
synchrotron loss (dashed line) and IC loss (dotted line). The results are
 for  the radial distance $r=0.1$~AU from the pulsar in opposite direction 
of the companion star  and the magnetic field $B=0.5$G. 
The turn over of the IC loss at the Lorentz factor $\sim 10^5$ 
is due to the Klein-Nishina effect. }
\label{radloss}
\end{center}
\end{figure}

We expect that the high-energy emission processes occure at the vicinity 
of the shock surface. Figure~\ref{radloss} shows the time scales of 
the radiation losses; solid, dashed and dotted lines show the time scales of 
the adiabatic loss, the synchrotron loss, and the IC loss, respectively. 
The results are for the radial distance $r=0.1$AU from the pulsar
 in opposite direction of the companion 
 and the magnetic field $B\sim0.5$~G. 
Since the time scale of the adiabatic loss ($\sim 3r/2v_{PW}$) 
represents the crossing time scale of the shock region,  
we can see in Figure~\ref{radloss} 
that the crossing time scale of  the particles with a  
Lorentz factor $\gamma_e>10^3$ is longer than the time scale
 of the radiation losses, implying 
the accelerated particles 
loose most of their energy at the vicinity 
of the shock surface through the radiation processes. 
In the present calculation, therefore, we take 
into account the emissions occurred between 
the shock distance $r_s$ and the  radial distance 
 $3r_s$, beyond which the emissions of the cooled  particles are 
negligible. 

Finally,  we expect that effect of  the Doppler 
boosting due to  the finite velocity of the shocked pulsar wind 
is the main reason to cause the temporal variation 
 of the X-ray emissions with the orbital phase. The Doppler boosting 
introduces an orbital modulation of the emissions that are isotropic in 
co-moving frame with the flow.  Dubus et al. (2010) suggested that 
the observed orbital modulation of the X-ray emissions from LS~5039 is
 the  result of the Doppler boosting of  the shocked pulsar  wind with a 
 mildly relativistic speed, $\beta_{PW}\equiv v_{PW}/c\sim 0.15-0.3$. 
Note that 
this scenario will be  different from the case of 
 the gamma-ray binary PSR B1259-63/LS2883. For PSR B1259-63/LS2883 system, 
which has a  highly eccentric orbit with $e=0.87$,  
the shock distance from the pulsar varies about a factor of ten along the 
orbital phase. This large variation in the shock distance can produce a large
 temporal variation of the synchrotron emissions from the shock 
(Tavani \& Arons 1997; Takata \& Taam 2009; Kong et al. 2011, 2012). 
With a moderate eccentricity ($e\sim 0.35$), on the other hand, 
the shock distance of LS~5039  system varies only about a factor of 
two along the orbital phase. The slightly change in  
the shock distance with the orbital phase will not be able to 
 reproduce the observed temporal variation in X-ray emissions
 from LS~5039.
\subsection{Pair-creation Process}
The high-energy TeV gamma-rays may be converted into the electron and 
positron pairs by colliding  with the soft-photons from the companion star. 
The mean free path of the pair-creation process of a photon with 
an energy $E_{\gamma}$ at a radial distance $\tilde{r}$ from the companion 
star is calculated from 
\begin{equation}
\frac{1}{\ell (\tilde{r}, E_{\gamma})}
=(1-\cos\theta_{\gamma\gamma})\int_{E_c}^{\infty}dE_s\sigma_{\gamma\gamma}dN_s/dE_s,
\end{equation}
where $dN_s/dE_s$ is the distribution of the number density of the  stellar soft photon, and 
\begin{equation}
\sigma_{\gamma\gamma}=\frac{3}{16}\sigma_{T}(1-v^2)
\left[(3-v^2){\rm \ln}\frac{1+v}{1-v}-2v(2-v^2)\right], 
\end{equation}
where $\sigma_T$ is the Thomson cross section, $v(E_\gamma, E_s)
=\sqrt{1-E_c/E_{\gamma}}$, and $E_c=2(m_ec^2)^2/[(1-\cos\theta_{\gamma\gamma})E_s]$
 with $\theta_{\gamma\gamma}$ being the collision angle. 

Since the  electron and positron pairs created by TeV gamma-rays have 
a Lorentz factor of $\sim 10^6$, they may emit  new gamma-rays
 via the inverse-Compton process. The gamma-rays emitted by the pairs 
in turn produce next generation of pairs. Hence, 
 the TeV gamma-rays emitted toward the companion star develop 
 the pair-creation cascade process.  
It has been suggested that the contribution of the emissions 
from new generation of the pairs will be important for the emissions around 
the SUPC phase, where the companion star locates between the pulsar and 
the observer (e.g. Sierpowska-Bartosik \& Torres 2007; 
Yamaguchi \& Takahara 2010; Cerutti et al. 2010). In the calculation,
 we assume that the created pair travels 
straight in  the direction of the momentum of the incident gamma-rays.

\subsection{Model Parameters}
\label{para} 
\begin{table}
\caption{Model fitting parameters. From top to bottom, spin 
down power of pulsar ($L_{sd}$), rotation period ($P$), surface magnetic 
field ($B_s$), typical Lorentz factor of the injected 
 cold-relativistic pulsar wind 
($\Gamma_o$), minimum Lorentz factor of the electrons/positrons accelerated 
at the shock ($\gamma_{min}$), index of the power law distribution ($p$), 
the velocity of the post shocked flow in units of 
speed of light ($\beta_{PW}$),  
 distance to the system ($d$), and inclination angle $i$, respectively.}
\label{table}
\begin{tabular}{clc}
\hline\hline
{\bf Pulsar} &  &  \\
$L_{sd}$ &  & $2\times 10^{36} {\rm erg~s^{-1}}$ \\
$P$ &  & 0.1s \\
$B_s$ &  & $2\times 10^{12}$G \\
\hline
{\bf Pulsar wind} &  &  \\
$\Gamma_0(r_i)$ &  & $5\times 10^3$ \\
\hline
{\bf Shocked pulsar wind} &  &  \\
$\gamma_{min}$ &  & $\Gamma_o(r_s)$ \\
$p$ &  & 2.1 \\
$\beta_{PW}$ & & 0.15 \\
\hline
{\bf System} &  &  \\
$d$ &  & 2.5kpc \\
$i$ &  & 65 degree \\
\hline\hline
\end{tabular}
\end{table}
In Table~\ref{table}, we summarize the parameters assumed in
 the model fitting.  The observed fluxes are explained by 
a spin down power $L_{sd}\sim 2\times 10^{36}{\rm erg~s^{-1}}$. With $L_{sd}\sim 2\times 10^{36}~\rm{erg~s^{-1}}$, if the pulsar has 
a typical magnitude of the surface  magnetic field, $B_s\sim 
2\times 10^{12}$G, the dipole radiation model of the pulsar spin down predicts 
 the rotation period of   $P\sim 0.1$~s, implying the  
 gap factional thickness (c.f. equation~(\ref{fgap})) and  
luminosity (c.f. equation~(\ref{lgamma})) of the gap 
emissions correspond to $f_{gap}\sim 0.3$  and 
$L_{\gamma}\sim  0.03L_{sd}$, respectively. 
The equation~(\ref{gamma0}) 
 implies that the initial Lorentz factor of the kinetically dominated flow 
is of order of $\Gamma_{0}=10^{4}$. In the present calculation, 
 we apply $\Gamma_{0}(r_i)=5\times 10^3$ to fit the data. 
The standard $Fermi$ first-order shock acceleration model has implied 
$p\sim 2$ as  typical  power law index of the distribution of
 the accelerated particles (e.g. Longair 1994 and references therein). 
The index $p\sim 2$  is also expected 
 from  typical photon index $\alpha_{X}\sim 1.5$ of the observed 
X-ray emissions from LS~5039. In the fitting, we will apply $p\sim 2.1$. 
 We adopt the flow velocity $\beta_{PW}=0.15$ 
 and the Earth viewing angle $i=65$~degree to reproduce  the amplitudes
 of the orbital modulation of the observed   X-ray and gamma-ray emissions.
 The distance 
to the system is assumed to be 2.5kpc.

\section{Comparison with the multi-wavelength observations}
\label{result}
\subsection{GeV emissions}
\label{gev}
The calculated phase-resolved spectra and the light curves are compared
 with the results of the \fermi\ in  Figures~\ref{fit} and~\ref{light-fit}, 
respectively.  We can see in the figures that the emissions from 
the outer gap (dashed lines), from 
cold-relativistic pulsar wind (dotted lines) and from the shocked pulsar 
wind (dashed-dotted lines) all contribute to the emissions observed 
by the \fermi. The emissions from the cold-relativistic pulsar 
winds contribute to the emissions around 0.1-0.5GeV and dominates
 other two components around 
the SUPC. The outer gap emissions dominate other two components in 1-10GeV 
energy bands for entire orbit. Above $\sim$10GeV, 
the emissions are dominated by the IC process of the shocked pulsar wind. 
We find that the calculated spectra and the light curves for the total 
emissions (solid lines) in the figures explain   
major properties of the observations; (1) the GeV flux becomes maximum around 
the SUPC and becomes minimum around INFC, (2) the amplitude of light curve 
for $>0.2$GeV bands is larger than that for $>1$GeV, and (3) there is a 
spectral cut-off at $\sim 2$GeV. The present model predicts that 
the emissions from the cold-relativist pulsar wind make the spectrum  
softer around the SUPC (c.f. Figure~\ref{light-ene}), which is also  
consistent with the observation. We found that within the framework of the calculations,  it would be  difficult to explain the position of the upper limit
 at $\sim20$GeV for the orbital phase $0.83<\phi<1.17$ 
(see section~\ref{upper}).

\subsection{Multi-wavelength emissions}
\label{multi}
\begin{figure}[t]
\begin{center}
\plotone{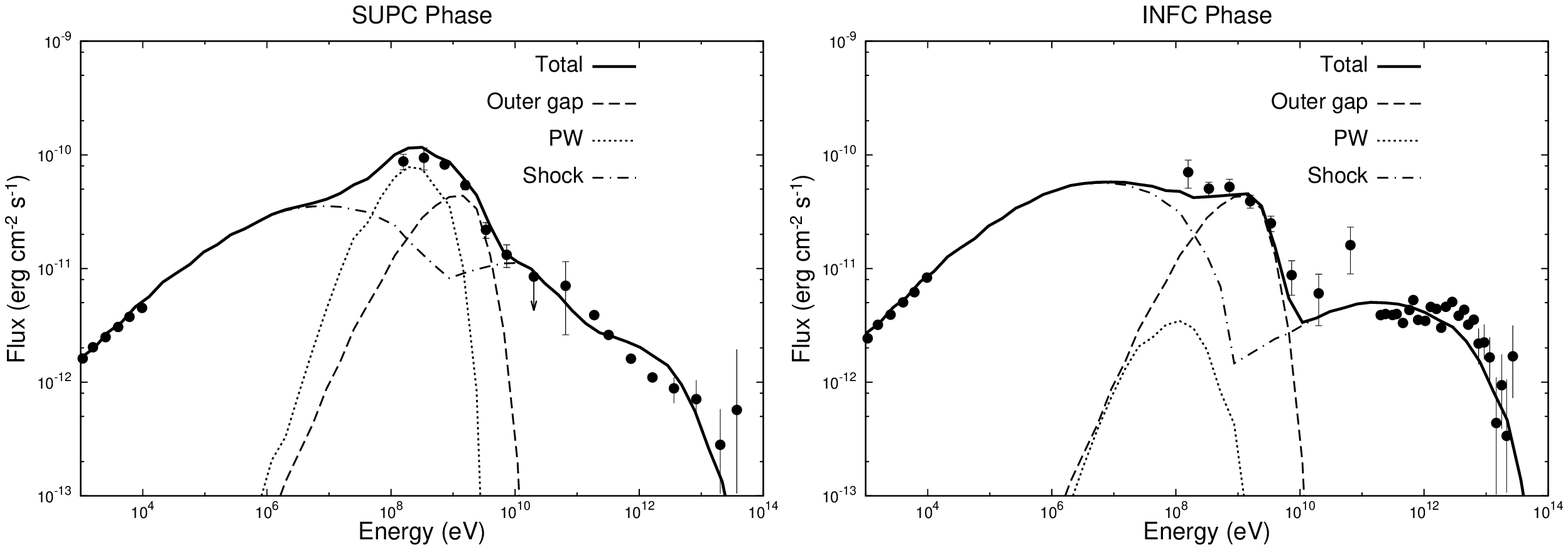}
\caption{Multi-wavelength spectra of LS~5039 averaged over SUPC phase, 
 $0.9<\phi<0.45$ (left), and  INFC phase, $0.45<\phi<0.9$ (right), respectively.
The dashed line, dotted line and dashed-dotted line represent calculated 
spectra of the emissions  from the outer gap, 
cold-relativistic pulsar, and shocked pulsar wind. The solid lines 
show the total emissions. The results are for $L_{sd}\sim 
2\times 10^{36}{\rm erg~s^{-1}}$
 $\sigma(r_{apex,0})\sim 0.2$, $\alpha=2$ and $p=2.1$.
The observation data are taken from Takahashi et al. (2009) 
for X-rays, current work for GeV  
and Aharonian et al. (2006) for TeV gamma-rays.}
\label{spec-fit}
\end{center}
\end{figure}

\begin{figure}
\begin{center}
\includegraphics[height=8cm,width=8cm]{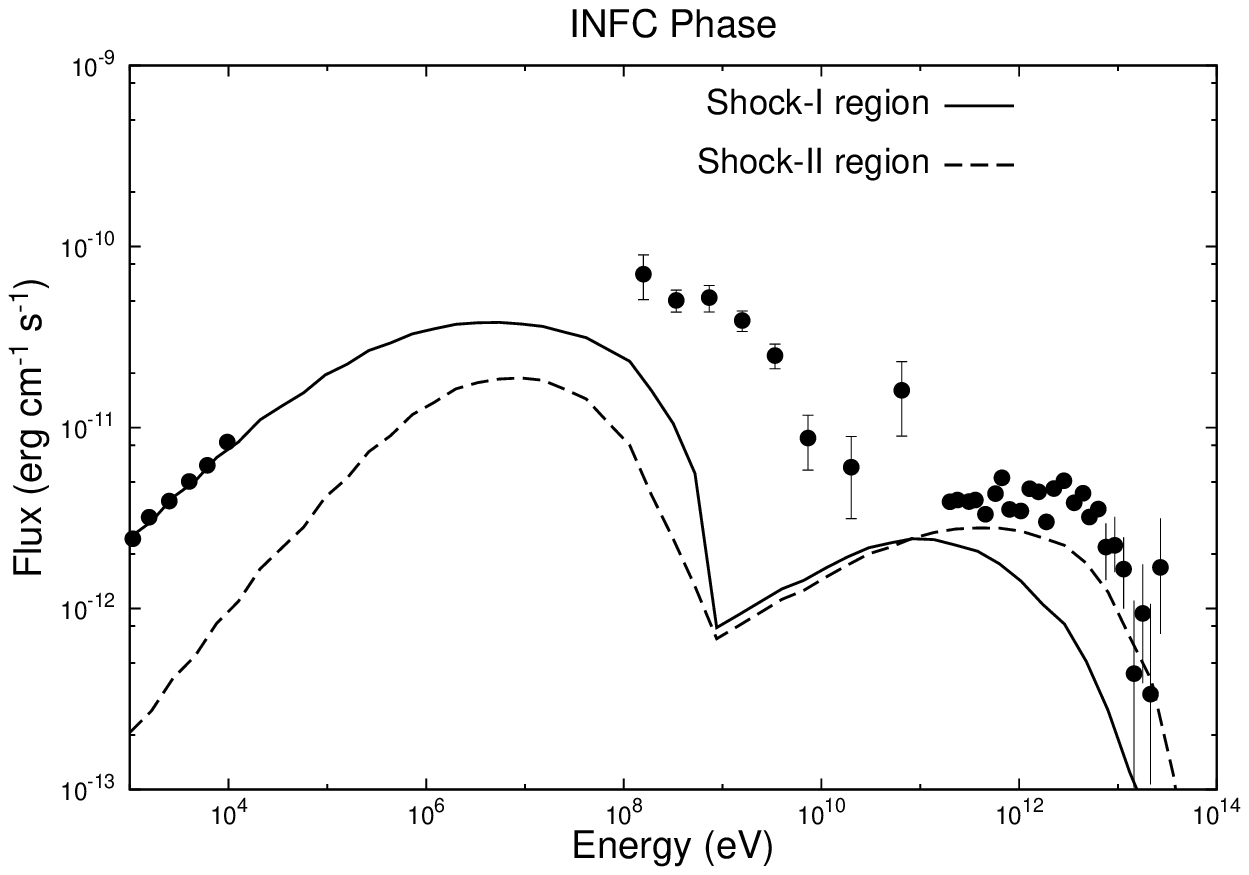}
\caption{Multi-wavelength spectra of the INFC phase. The solid line 
and dashed line show the calculated spectra of the particles 
 accelerated at Shock-I region and Shock-II region, respectively. 
 The emissions from the magnetosphere and the 
cold-relativistic pulsar wind are not displayed in the figure. }
\label{spec-shock}
\end{center}
\end{figure}

Figure~\ref{spec-fit} compares the calculated spectra with 
the multi-wavelength observations. In the figures,  
the dashed line, dotted line and dashed-dotted line 
represent the calculated spectra of the curvature emission in 
 the outer gap, of the IC process of 
cold-relativistic pulsar, and of the shock 
 emissions (synchrotron below $\sim100$MeV and 
IC above $\sim 100$MeV), respectively. 
The solid lines show the spectra combining  each component.  
In the calculation, we assumed that the magnetization parameter at 
the shock decreases with
 the inverse square of the shock distance, that is, 
  $\sigma (r_s)=0.2(r_s/r_{apex,0})^{-2}$. 

\subsubsection{Shock-I v.s. Shock-II}

 One important prediction in the present scenario 
is that the X-rays and TeV 
gamma-rays are originated from different shock regions. We have 
assumed that there are two kind of shocks, that is,  one (Shock-I)  is 
located position at which  the pulsar wind  pressure and stellar wind pressure 
are in balance, and other (Shock-II) is located at $r_s\sim D_{sep}$ 
in opposite direction of the companion star 
(see Figure~\ref{LS5039}). Figure~\ref{spec-shock} 
compares the respective contributions of the emissions from Shock-I
 (solid line) and from Shock-II (dashed line) to the total emissions.  
 We find in the figure  that the X-ray emissions and TeV emissions are 
mainly produced by Shock-I and Shock-II regions, respectively.
 
The difference in the spectral properties of the Shock-I and Shock-II 
are caused by the difference in the assumed  magnetic field strength 
at each region. In calculation, we assumed that the magnetization 
parameter develops as 
$\sigma (r_s)=0.2(r_s/r_{apex,0})^{-2}$, which produces 
$B\sim 15$G and $B\sim 0.5$G as the magnetic field strength of 
the Shock-I and Shock-II, respectively.  As we can see 
in Figure~\ref{spec-shock}, the synchrotron radiation 
of Shock-I is stronger than that of Shock-II. In  Shock-I region, 
the synchrotron cooling time scale  
for the particles with a Lorentz factor $\gamma_e>10^6$ is shorter than  
 the  IC cooling time scale, and as a result 
 the spectrum of IC does not extend beyond $\sim 0.5$TeV.
 With $B\sim 0.5$G in the Shock-II region, on the other hand, 
 the IC cooling dominates the synchrotron cooling for the electrons/positrons 
 with a Lorentz factor up to  $\gamma_e\sim 10^{7}$, and hence 
the spectrum of IC emissions can extend  to $\sim 5$TeV.

 The result that TeV emissions are produced by Shock-II region with 
a magnetic field of $\sim 0.5$G is  consistent 
with the results obtained by Zabalza et al. (2013). We showed that 
X-ray emissions in Shock-II region could not explain the X-ray 
observation, which is also consistent with the conclusion of 
Zabalza et al. (2013). Hence, we propose that  the 
 particles accelerated at  the Shock-I  produces the X-rays via 
synchrotron radiation, while  Zabalza et al. (2013) assumed that the 
shocked particles produce the Gev emissions via the IC process.  As we 
discussed in section~\ref{gev}, our model expected that 
the 0.1-10GeV emissions are composed of the magnetospheric emissions 
and the pulsar wind emissions.

\subsubsection{Orbital variations}
\begin{figure}
\begin{center}
\includegraphics[height=10cm]{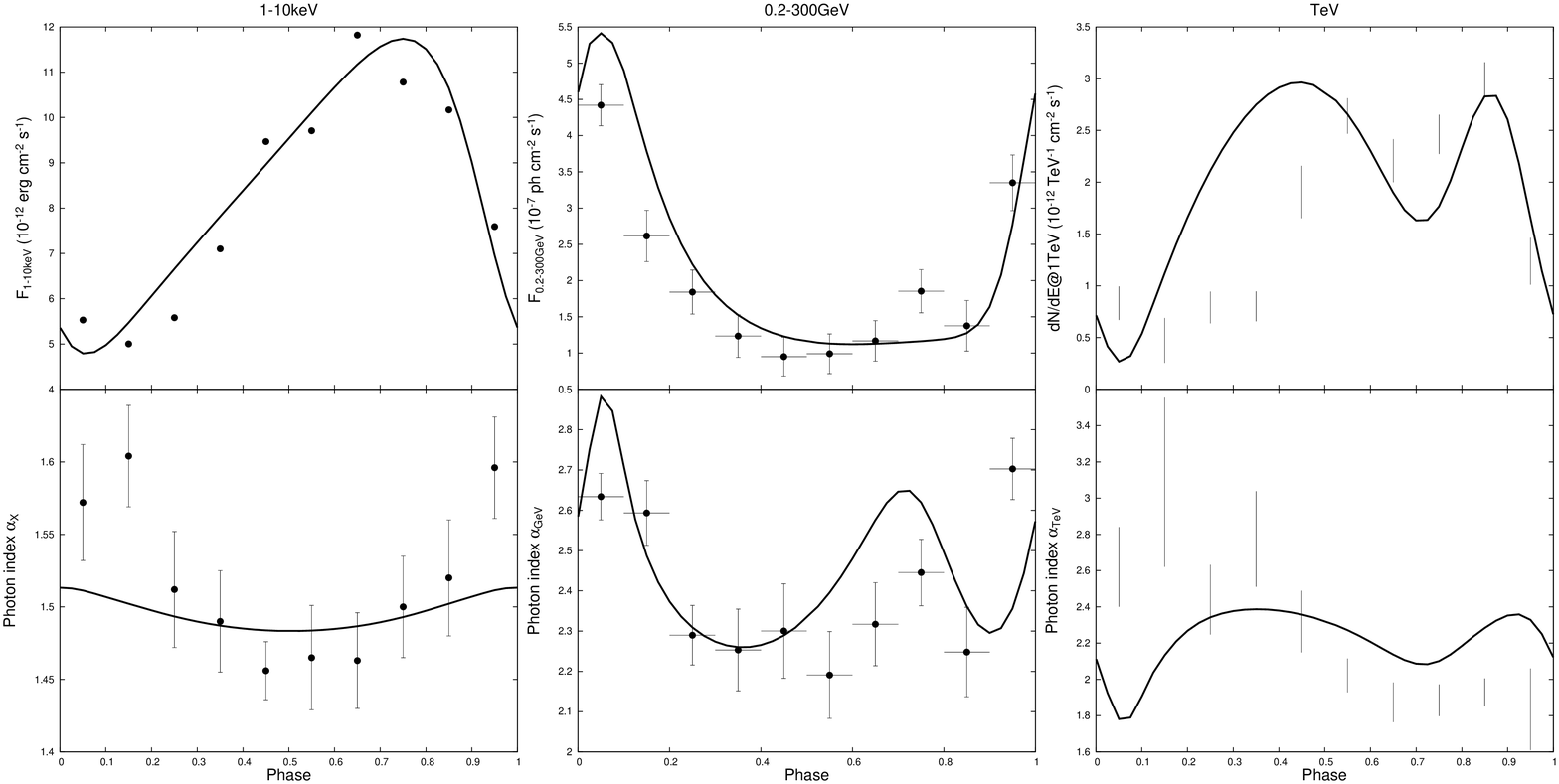}
\caption{Orbital variation of the flux and photon index 
in X-ray (left panel), GeV (middle panel) and TeV (right panel), respectively.
X-ray data and TeV data were taken from Takahashi et al. (2009) and Aharonian et al. (2009), respectively. The GeV data show the result of 
the present work (section~\ref{observation}). 
The lines show the results of the emission model 
discussed in section~\ref{model}.}
\label{light-ene}
\end{center}
\end{figure}
Figure~\ref{light-ene} summarizes the orbital variations of the flux (top panels) and photon index fitted by a single power law function (bottom panels)
 for 1-10keV (left), 0.2-300GeV (middle) and 0.2-5TeV (right), respectively.  
For X-ray emissions (left panel), we find that  the Doppler boosting 
with  $\beta_{PW}=0.15$ of the post-shock flow velocity can reproduce
 the observed amplitude in X-ray band.  This result is consistent
with that of Dubus et al. (2010). With $p=2.1$ for
the power law index of the  particle distribution at the shock,
the predicted photon index $\alpha_{X}\sim 1.5$ is qualitatively consistent
with the result of the observation. 

For 0.1-1 GeV energy bands, the cold-relativistic
 pulsar wind contributes
 to the calculated emissions, and therefore the orbital modulation of 
the flux shows  different  behavior  from what the X-ray emissions show, 
as we can seen in Figure~\ref{light-ene}. 
The cold-relativistic pulsar wind 
mainly produces  0.2-0.5GeV gamma-rays and its emissions  dominate the 
outer gap/shock emissions around  SUPC (see Figure~\ref{fit}).  
Hence, the calculated light curve in near GeV energy has a flux peak  at  around the SUPC.  
This model predicts that the GeV spectrum is softer around SUPC and harder around apastron.
 We also find in the Figure~\ref{light-ene} that the GeV spectrum locally becomes soft
 around the INFC. This is because the inverse-Compton process of the shocked pulsar wind 
is less efficient around INFC (c.f. the right panel of Figure~\ref{light-ene}) 
  and as a result the flux above 10~GeV tends to decrease 
around INFC.  
We can see in Figure~\ref{light-ene} that the properties of the calculated 
orbital modulation of  GeV gamma-rays are consistent with the observations.

For TeV energy bands (right panel), we find 
that the calculated light curve shows double peak structure  around the INFC,
which could be consistent with the observations.
Since  most of the photons emitted around
SUPC cannot escape from the pair-creation process,
 the TeV flux   tends to increase as the pulsar moves toward the INFC.
As the pulsar approaches to  INFC, since the
 IC process with the  tail-on like collision produces  the
TeV photons traveling toward the Earth, the radiation efficiency
 decreases. In the calculated light curve, therefore,
a dip appears  around the INFC. 

 The present model will overestimate a  TeV flux 
at $0.1-0.5$ orbital phase, as the right-upper panel 
in Figure~\ref{light-ene} shows.  
Since the TeV gamma-rays are mainly produced at the Shock-II region (c.f. Figure~\ref{spec-shock}), 
this discrepancy may suggest that 3D geometry of the Shock-II region 
is more complex than one assumed in this study, for 
 which the result of 2D simulation has been 
applied. The detailed analysis with the shock geometry  obtained 
by 3D simulation will be worth investigating further.

We also find in the bottom panel 
that difference between the observed and predicted  photon indexes 
at 0.2-5TeV energy band is suggestively large around SUPC; 
the calculated spectrum around SUPC becomes very hard compared with 
the observed spectrum. The predicted hard spectrum at 0.2-5TeV energy
 band is caused by the effect of the pair-creation process.
Since the optical depth around 0.1TeV is larger than that around 1TeV,
the pair-creation process absorbs  0.1TeV photons more than 1TeV photons,
 and hence the spectrum at 0.2-5TeV
tends to have a photon index smaller than two.
The large difference in photon indexes may suggest that an additional
 component, for example nebula component (Bednarek \& Sitarek 2013), 
contributes to  the observed TeV  emissions around the SUPC.

\section{Discussion}
\label{discussion}
\begin{figure}
\begin{center}
\includegraphics[height=8cm,width=8cm]{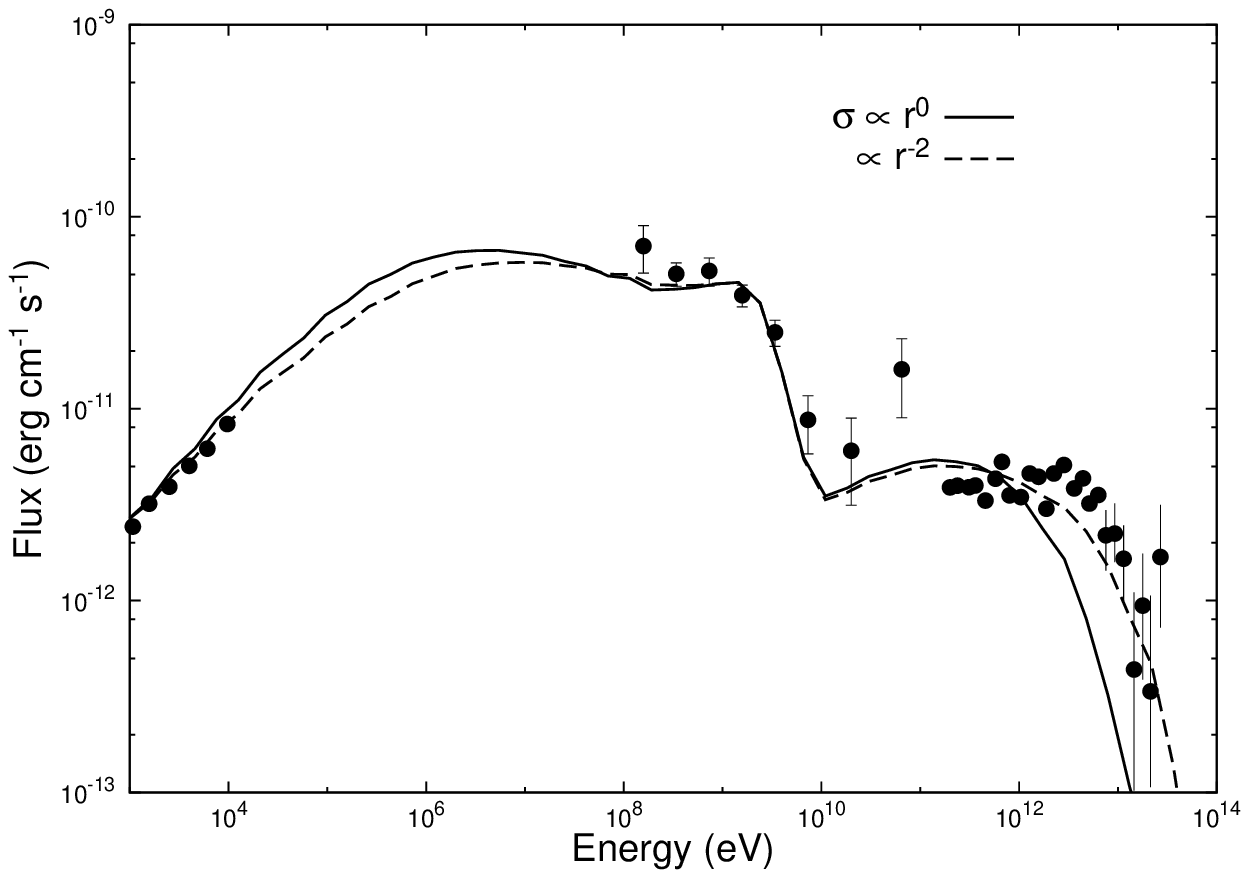}
\caption{Multi-wavelength spectra of the INFC phase. The different lines 
show the calculated spectra for different power index $\alpha$ in the 
radial distribution of the magnetization 
parameters $\sigma(r_s)\propto r_{s}^{-\alpha}$. The results are 
for  $\alpha=0$ (solid line) and 2 (dashed line).}
\label{spec-sigma}
\end{center}
\end{figure}

\begin{figure}
\begin{center}
\includegraphics[scale=0.5]{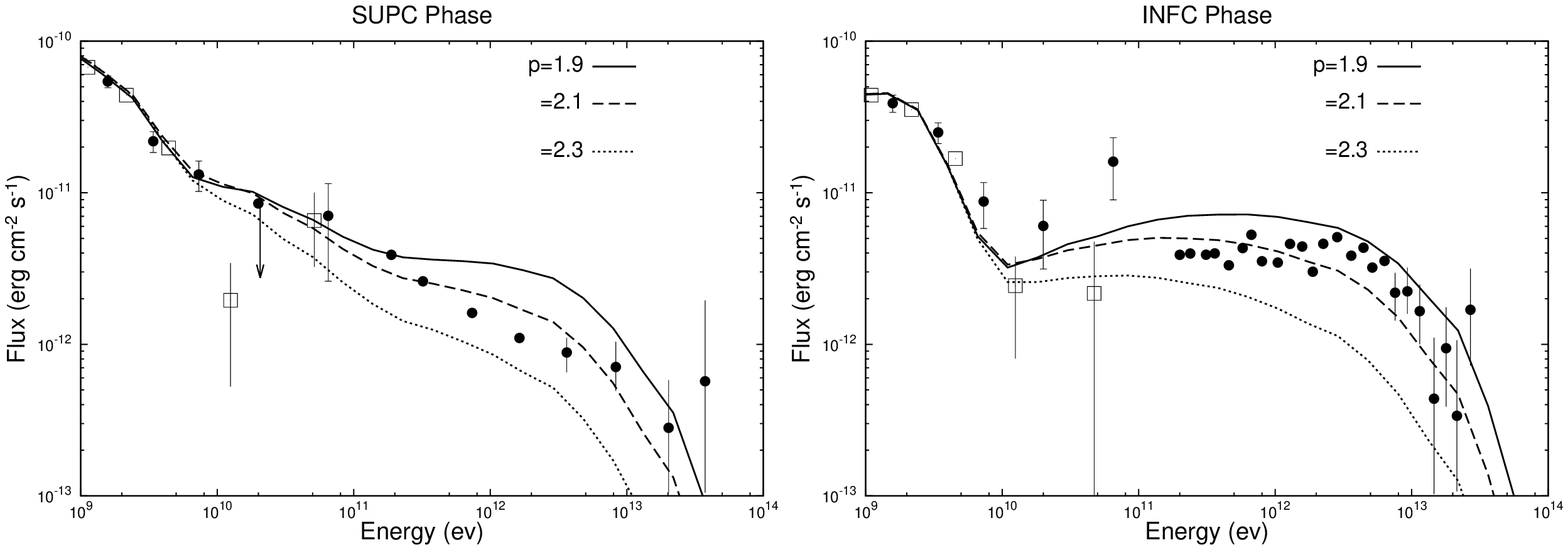}
\caption{GeV/TeV spectra averaged over the SUPC phase (left) and 
INFC phase (right). The solid, dashed and dotted lines represent the results 
for power law index of $p=1.9$, 2.1, and 2.3, respectively.
 The boxes show the data taken from Hadash et al. 2012.}
\label{tevcomp}
\end{center}
\end{figure}

\begin{figure}
\begin{center}
\includegraphics[scale=0.5]{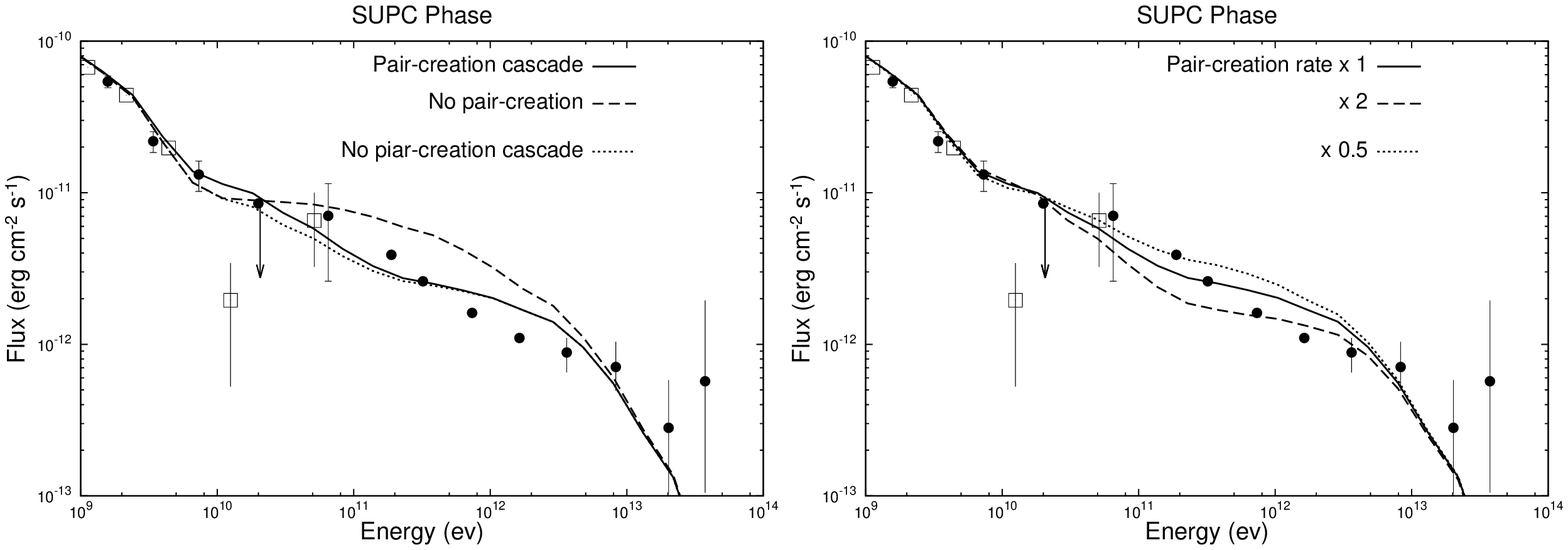}
\caption{GeV/TeV spectra averaged over the SUPC phase. In the 
left panel, different lines correspond the 
calculated spectra with different effects of the  pair-creation 
process. The solid line shows the spectrum taking account
the pair-creation cascade. The dotted line takes into account the pair-creation
process, but it ignores the emissions from the created pairs and subsequent
 pair-creation cascade.
The dashed line represents the spectrum, for which the pair-creation 
process is ignored. In the right panel, the different lines 
represent the results for the different pair-creation rate. The dashed and 
dotted lines show the spectra for the pair-creation rate increased  
and decreased by a factor of 2, respectively. 
The boxes show the data taken from Hadash et al. 2012.}
\label{pairc}
\end{center}
\end{figure}

\subsection{Dependency on magnetization parameter}
\label{depmag}
The calculated spectrum in TeV energy bands in fact depends 
on how the magnetization parameter at the shock evolves with 
the shock distance from the pulsar, that is, 
the power index $\alpha$. 
 Figure~\ref{spec-sigma} compares the calculated spectra of INFC phase 
with index  $\alpha=0$ (solid line) and 2 (dashed line).
 As we can see  in Figure~\ref{spec-sigma}, the calculated spectrum with  
the constant magnetization parameter ($\alpha=0$) predicts a cut-off energy in 
TeV bands much smaller than the observations.
 Since a smaller index $\alpha$ predicts  a larger magnetic field 
and hence a larger synchrotron cooling in Shock-II region, the TeV emissions 
are suppressed. In the present model, therefore,  
a larger  power index $\alpha$  is preferable in explaining  the 
 observed X-ray and TeV emissions, simultaneously.  

\subsection{Dependency on the particle distribution}
 Figure~\ref{tevcomp} summarizes the dependency of the TeV spectra on 
the power law index of the energy distributions of the particles 
at the shock; the left and right panels  show the spectra 
for SUPC phase and INFC 
phase, respectively. As the SUPC phase, we can see that the calculated 
spectrum with the index $p=1.9$ (dashed line) will be too hard compared with 
the observed results, while the calculated spectrum with $p=2.3$ 
(dotted line) is too soft. Within the framework of current model,
 therefore,  the index $p\sim 2.1$  (solid line) provides a better 
fit for the observed TeV spectra of SUPC.
  We also note that the observed index 
$\alpha_X=1.5\sim 1.6$ of the X-ray emissions are fitted better by 
$p\sim 2.1$.

\subsection{Dependency on parameter $\eta$}
The ratio ($\eta$) of the momenta of the pulsar wind and the stellar wind 
is also model parameter. By assuming  $L_{sd}\sim 2\times 10^{36}{\rm erg~s^{-1}}$, which can explain 
the observed flux with $d=2.5$kpc,  and using typical mass loss 
rate $\dot{M}\sim 10^{-7}M_{\odot}{\rm yr^{-1}}$ of O-type main sequence star, we have applied 
the ratio $\eta=0.05$ in the present calculation. With the present calculation, 
it is difficult to constrain the reasonable range of possible $\eta$ by fitting of 
the observational results. We have used the momentum ratio to determine 
the distance to the shock apex from the pulsar. The magnetic field 
strength at the shock is an important quantity to determine the 
properties of the calculated 
spectra. In the present study, however, since the magnetization parameter 
and hence the magnetic field strength at the shock are also model parameters, 
we can adjust the magnetic field strength for each $\eta$ to fit the observed 
spectra. The reasonable range of 
$\eta$ cannot be constrained by fitting the observed spectra.
 
A study of the orbital modulation cound constrain 
the range of $\eta$, since the 
geometry of the shock (e.g. opening angle) depends on $\eta$. 
The calculation with the Doppler 
effects and 3D geometry of the shock will 
produce the different properties of the orbital modulation for different $\eta$. In the present 
calculation, however, we ignored the effect of 3D geometry, when we calculated the orbital modulation. 
By assuming that the flow of shocked pulsar wind points radially outward from the companion star,
 we took into account only the effect of the Doppler boosting. 
In such a case, we cannot reasonably 
constrain the possible $\eta$ by fitting the orbital modulation. 
The full calculation with 3D geometry will be subject to the future study.

\subsection{Effect of the pair-creation cascade}
Figure~\ref{pairc} shows the GeV/TeV spectrum of the SUPC phase and 
compares the  calculated spectra with different type of consideration on 
the pair-creation process. In the left panel, 
the solid line represents the spectrum 
including  the effects of the pair-creation cascade. 
The dotted line takes into account the absorption by the
pair-creation process, but it ignores
 the emissions from the created pairs. The dashed line shows the spectrum 
ignoring the pair-creation process. For 
LS~5039, because the surface temperature of the companion 
star is $kT\sim 3$eV, the gamma-rays with an energy 0.05-5TeV are subject to 
 the pair-creation process, as Figure~\ref{pairc} shows. The emissions 
from the new pairs and the subsequent pair-creation cascade processes affect
 the spectra at $\sim$10GeV energy band, as the  dotted line of 
Figure~\ref{pairc} shows .

 In the right pane, the different 
lines show the calculated TeV spectra
 with different optical depth of the pair-creation. 
 The solid line shows the result for the optical depth that assumes
 $kT_s=3.4$keV and the spherically symmetric stellar photon field. 
There will be several uncertainties related to the 
stellar photon field at the emission region; (1) 
 the stellar photons field could depend on the
 latitude (Negueruela et al 2011), (2) the spectrum  would not be exactly 
described by  Planck function, and (3) the photon density at the emission region 
will depend the complex shock structure. To see the dependency 
of the spectral shape in TeV energy bands, we artificially increased or 
decreased the optical depth of the pair-creation process. 
 For  the dashed  line in the right panel of Figure~\ref{pairc}, 
we  increased the optical depth by factor of two,  while 
 for the dotted line we decreased it by the factor of 2. 
As we can see in the figure, the difference in the optical depth affects to 
the spectrum in 0.1-1TeV energy bands.

\subsection{10-100GeV emissions}
\label{upper}
 Although the present model can explain many  observational properties in 
the multi-wavelength bands, it is unsure that the present model is consistent 
with the spectral behavior of the 10-100GeV emissions at SUPC phase 
observed by \fermi. As Figure~\ref{tevcomp} shows, 
 the calculated spectrum in SUPC does not show a spectral break in 10-100GeV 
bands, while upper limit around $\sim20$GeV determined by 
 the $Fermi$  may suggest the existing of a spectral break.
 To explain the position of the upper limit,  
one may consider that the minimum Lorentz factor of the 
shocked particles is of order of $\gamma_{e,min}\sim 10^5$. 
 In the present model, since we have expected
$\Gamma_{0}\sim 5\times 10^3$ for the typical  Lorentz factor of the cold-relativistic
pulsar wind,  we have assumed that the shocked particles have a 
Lorentz factor larger than  $\gamma_{e,min}\sim 5\times 10^3$. Furthermore, 
we can see that if the minimum Lorentz factor of the shocked particles is 
 larger than  $\gamma_{e,min}\sim 10^5$, the predicted spectrum of 
X-ray emissions becomes  much harder  
than results of the observation, which shows a photon idnex $\alpha_X\sim 1.5$. 
This hard spectrum in 0.1-10keV bands is expected,  because the energy of 
the synchrotron photons emitted by the 
 particles with a Lorentz factor $\gamma_{e,min}\sim 10^5$ is larger than 10keV. 
 To investigate the behaviors of emissions in 10-100GeV bands of the SUPC, 
 more detailed theoretical and observational studies would be required. 

\section{Summary}
\label{summary}
 In this paper, we have discussed the mechanisms of the high-energy emissions 
from the  gamma-ray binary LS~5039. In the first part, 
we reported on results of the  observational analysis using 
 four year data of \fermi\ and updated the information of the GeV emissions 
from LS~5039.
 We showed that due to the improvement of instrumental response function 
 and increase of the statistics,  the flux in $\sim$100MeV bands  
has noticeably decreased and  uncertainties of the 
spectra beyond $\sim10$GeV have been significantly improved. 
We divided the observation time into three equally spaced orbital phase bins, 
for which one bin includes the emissions from superior conjunction. We showed 
 that the spectra of two orbital bins  excluding  the superior conjunction 
have a clear spectral cut-off at several GeV  and they 
resemble to those of the gamma-ray pulsars. For the bin including 
the superior conjunction,  an enhancement at $0.1-0.3$ GeV
 is exclusively seen and the spectrum below 10GeV 
is significantly softer compared with the spectra of other two orbital bins, 
suggesting an additional component below 1GeV.
 Our results suggest that the 0.1-100GeV emissions 
from LS~5039 contain three different components, that is, 
 (i) the first  component contributing  to $<$1GeV emissions 
around superior conjunction,
 (ii) the second component dominating in the 1-10GeV emissions 
for entire component, and 
(iii) the thirst component which is compatible to lower energy 
tail of the TeV emissions

  In the second part, we discussed the emission mechanisms  
 of X-ray, GeV and TeV gamma-rays.  
We developed  the model, in which  the curvature 
emissions from the magnetospheric outer gap  
and IC process of
 the cold-relativistic pulsar wind contribute to the observed 
GeV emissions. Our model predicts that  the outer gap emissions mainly produce
 the observed emissions in 1-10GeV bands for entire orbit and 
the observed emissions near 1GeV are pulsed.  The 
IC process of the  cold-relativistic pulsar 
wind produces the gamma-rays with 0.1-0.5 GeV, and contributes to 
the observed spectrum at  SUPC phase.  
We applied the shock geometry resembles to that in Zabalza et al. (2013), 
that is, there are two kinds of  termination 
shock around  pulsar; Shock-I due to the pulsar wind/stellar wind 
 interaction   and Shock-II caused by 
 the effect of the orbital motion. We proposed that TeV gamma-rays 
are produced via the IC process of the Shock-II region, where the 
magnetic field strength is $\sim 0.5$G. This result 
on the emission region of the TeV gamma-rays is consistent with the result 
obtained by Zabalza et al. (2013). However, our model expects 
that the particles accelerated at  the Shock-I  produce the X-rays via 
synchrotron radiation, while  Zabalza et al. (2013) assumed that a strong 
radiative loss limits the acceleration at Shock-I and the shocked particles 
produce the Gev emissions via the IC process.

\acknowledgments
JT and KSC are supported by a GRF grant of HK
Government under HKU7009 11P. AKHK is supported
by the National Science Council of the Republic of China
(Taiwan) through grant NSC100-2628-M-007-002-MY3
and NSC100-2923-M-007-001-MY3. PHT is supported
by the National Science Council of the Republic of China
(Taiwan) through grant NSC101-2112-M-007-022-MY3.
CYH is supported by the National Research Foundation
of Korea through grant 2011-0023383. 
JT  thanks the  Theoretical Institute
for Advanced Research in Astrophysics (TIARA) operated under the Academia 
Sinica Institute of Astronomy and Astrophysics, Taiwan,
which  enable author (J.T.) to use the PC cluster at TIARA.





\clearpage

\end{document}